\documentclass[11pt,preprint]{aastex}

\usepackage{apjfonts}
\usepackage{float}
\usepackage{rotating}
\usepackage{epsfig}
\usepackage{lscape,graphicx}
\usepackage{url}

\defcitealias{cmg+06}{CD06}
\defcitealias{cgm+09a}{CD09}

\shorttitle{Origin of the Virgo Stellar Substructure}
\shortauthors{Carlin et al.}

\begin{document}

\title{The Origin of the Virgo Stellar Substructure}

\author{
Jeffrey L. Carlin\altaffilmark{1,4},
William Yam\altaffilmark{1},
Dana I. Casetti-Dinescu\altaffilmark{2},
Benjamin A. Willett\altaffilmark{1}, 
Heidi J. Newberg\altaffilmark{1},
Steven R. Majewski\altaffilmark{3}, \& 
Terrence M.Girard\altaffilmark{2}
}

\altaffiltext{1}{Department of Physics, Applied Physics, and Astronomy, Rensselaer Polytechnic Institute, 110 8th Street, Troy, NY 12180, USA (carlij@rpi.edu)}
\altaffiltext{2}{Astronomy Department, Yale University, P.O. Box 208101,
New Haven, CT 06520-8101}
\altaffiltext{3}{Department of Astronomy, University of Virginia,  P.O Box 400325, Charlottesville, VA 22904-4325}
\altaffiltext{4}{Visiting Astronomer, Kitt Peak National Observatory and Cerro Tololo Inter-American Observatory, National Optical Astronomy Observatory, which are operated by the Association of Universities for Research in Astronomy (AURA) under cooperative agreement with the National Science Foundation.}

\begin{abstract}

We present three-dimensional space velocities of stars selected to be
consistent with membership in the Virgo stellar
substructure. Candidates were selected from SA~103, a single
$40\arcmin \times 40\arcmin$ field from our proper motion (PM) survey
in Kapteyn's Selected Areas (SAs), based on the PMs, SDSS photometry,
and follow-up spectroscopy of 215 stars. The signature of the Virgo
substructure is clear in the SDSS color-magnitude diagram (CMD)
centered on SA~103, and 16 stars are identified that have high
Galactocentric-frame radial velocities ($V_{\rm GSR} > 50$ km
s$^{-1}$) and lie near the CMD locus of Virgo. The implied distance to
the Virgo substructure from the candidates is $14\pm3$ kpc. We derive
mean kinematics from these 16 stars, finding a radial velocity $V_{\rm
GSR} = 153\pm22$ km s$^{-1}$ and proper motions $(\mu_\alpha \cos
\delta, \mu_\delta) = (-5.24, -0.91)\pm(0.43, 0.46)$ mas
yr$^{-1}$. From the mean kinematics of these members, we determine
that the Virgo progenitor was on an eccentric ($e \sim 0.8$) orbit
that recently passed near the Galactic center (pericentric distance
$R_p \sim 6$ kpc). This destructive orbit is consistent with the idea
that the substructure(s) in Virgo originated in the tidal disruption
of a Milky Way satellite. $N$-body simulations suggest that the entire
cloud-like Virgo substructure (encompassing the ``Virgo Overdensity''
and the ``Virgo Stellar Stream'') is likely the tidal debris remnant
from a recently-disrupted massive ($\sim 10^9 M_{\sun}$) dwarf
galaxy. The model also suggests that some other known stellar
overdensities in the Milky Way halo (e.g., the Pisces Overdensity and
debris near NGC~2419 and SEGUE~1) are explained by the disruption of
the Virgo progenitor.

\end{abstract}

\keywords{Galaxy:structure -- Galaxy: kinematics and dynamics -- Galaxy: stellar content -- Proper motions -- stars: kinematics and dynamics -- galaxies: dwarf}

\section{Introduction}

Galaxy formation is currently thought to proceed via the hierarchical
merging of numerous smaller substructures, which are assimilated into
the host galaxies by tidal disruption and, ultimately, dissolved
(e.g., \citealt{j98a, ans+03a, bj05, fjb+06a}). This is by no means a
new idea, as it was suggested by
\citet{sz78} that late infall after the collapse of the initial
protogalaxy was responsible for the apparent age spread among Milky
Way (MW) globular clusters, and later (by, e.g., \citealt{m93, mmh96})
that the build-up of the Galactic halo proceeds via agglomeration of
dwarf galaxies. We have now seen this process in action via the
numerous stellar tidal streams that have been discovered coursing
through the MW's halo (see \citealt{g10} for a summary) and various
other stellar substructures in the form of ``clouds'' of stars (the
properties of known structures are summarized in \citealt{r10a}). In
the era of large-area, deep photometric and spectroscopic surveys, we
are now entering the regime where not only narrow, kinematically-cold
stellar streams can be found, but also the clouds and shells of debris
seen in, e.g., the \citet{jbs+08} models of hierarchical formation of
Milky Way-like galaxies. The menagerie of morphological structures
predicted by these models are now being seen with deep imaging around
external galaxies (e.g., \citealt{cmg+11, mgc+10}). Because of our
perspective within the Galaxy, such features are more difficult to see
in context and characterize. However, a look at the variety of
morphological types of known structures discussed in \citet{r10a}
shows that large photometric surveys have begun finding the predicted
remnants of halo assembly. The kinematical characterization of the
clouds and shells that is needed to assess their origins and discern
the nature of their progenitors has been more elusive, since
identification of stellar members of such low surface-brightness
features is difficult, and derivation of proper motions for such
distant structures is challenging. The particular one of these
apparent cloud-like stellar structures upon which we focus this work
is the overdensity (or, perhaps, collection of stellar excesses) in
the constellation Virgo.

The first recognition of a stellar overdensity in the halo near the
Virgo constellation was the finding by \citet{vza+01} of a cluster of
5 RR Lyrae stars within $195\arcdeg<$ RA
$\lesssim200\arcdeg$, at a distance of $\sim$20 kpc in the first 100
deg$^2$ results from the QUEST survey. Shortly thereafter,
\citet{nyr+02} used F-type turnoff starcounts on the
celestial equator to identify stellar overdensities, including one at
$180\arcdeg<$ RA $<195\arcdeg$, Dec $=0\arcdeg$ that was dubbed
S297+63-20.0 (the naming scheme denotes that the feature is at
$(l,b)\sim(297\arcdeg,+63\arcdeg)$ with turnoff magnitude
$g\sim20$). This feature, which is near the same region of sky as the
\citet{vza+01} RR Lyrae, shows a clear main sequence extending
downward from $g\sim19$, corresponding to a distance (assuming an old,
metal-poor population) of $\sim20$ kpc.  Since the distance is also
similar to that of the \citet{vza+01} stars, this large-area
overdensity might be associated with the RR Lyrae. \citet{nyr+02} also
noted that the color of the turnoff stars in S297+63-20.0 is
intermediate between that of Sagittarius tidal debris and the typical
MW halo turnoff, suggesting that the Virgo feature is related to
neither of those structures.

Further results from the QUEST survey, which studied a 2.3-degree wide
strip centered at Dec $\sim -1\arcdeg$, were presented in
\citet{vz03}. In the 380 deg$^2$ portion of the QUEST survey
discussed, the authors find 497 RR Lyrae between 4-50 kpc. Many of the
distant RR Lyrae are Sgr debris, but the authors identify a distinct
clump at mean magnitude of $\langle V_0 \rangle =16.9\pm0.2$,
corresponding to a distance of $\sim19$ kpc. Because of its location
at $\langle RA \rangle \sim186\arcdeg$, this grouping was dubbed the
``12.4-hour clump''. Based on the RR Lyrae periods and
period-amplitude distributions, it appears that the clump contains RR
Lyrae of both Oosterhoff classes, making them more typical of dwarf
galaxy populations than globular clusters. This 12.4-hour clump is
within the same region and at the same distance as S297+63-20.0, which
suggests an association. A ``fluff'' of overdensity was also seen in
the \citet{msw+03} study of 2MASS-selected M-giant stars at the same
position ($150\arcdeg<RA<210\arcdeg$) as S297+63-20.0 and the
12.4-hour clump, though it was suggested by those authors to be part
of the ``descending, foreshortened loop'' of the Sagittarius leading
tidal tail. This idea was explored further by \citet{mga+04}, who
developed a model of the Sgr disruption that produced debris at
roughly the distance, position, and velocity of the \citet{vz03} RR
Lyrae and the \citet{nyr+02} main sequence detections.

The first spectroscopic study of any of these overdensities came when
\citet{dzv+06} observed 18 RR Lyrae from the QUEST survey. Their study
found that a clump of stars centered at $\alpha\sim186\arcdeg
(\pm5\arcdeg)$ and $R_{\rm Sun}=19.6$ kpc is also clustered in
velocity. The six stars making up this clump have a mean $\langle
V_{\rm GSR} \rangle =99.8$ km s$^{-1}$, and a dispersion of 17.3 km
s$^{-1}$. This clump of stars was dubbed the ``Virgo Stellar Stream
(VSS)''. Duffau et al. then expanded the sample to a larger area and
RVs within $\sim3\sigma$ of the peak, and identified 13 stars with
$\langle V_{\rm GSR} \rangle =85$ km s$^{-1}$ with $\sim25-30$ km
s$^{-1}$ dispersion. These stars are spread over a distance range from
$R_{\rm Sun} = 16-20$ kpc (with one at $\sim24$ kpc); thus, the VSS
must be large and diffuse (or contain multiple overdensities).  The
metallicity of the 7 RRab stars selected to be members is $\langle
[$Fe/H$] \rangle =-1.86$, with $\sigma_{\rm [Fe/H]} = 0.4$. The large
spread in conjunction with the lack of an obvious intermediate-age
population in the SDSS CMDs is taken as evidence of a low-luminosity,
old, metal-poor dwarf galaxy progenitor for the VSS.

Further detections of halo overdensities from the QUEST RR Lyrae
survey's first catalog were presented by \citet{vz06}, who found that
after removing the contribution of their best-fitting smooth halo
model, the Virgo structure is visible as a peak at RA$\sim186\arcdeg$
at a distance of $\sim17$ kpc (spanning 12-20 kpc between
$175\arcdeg<$RA$<205\arcdeg$). Velocity substructures were seen among
bright ($V<16.1$) QUEST RR Lyrae by \citet{vjz+08}, including a
grouping at $10.5<D<12$ kpc with $\langle V_{\rm GSR} \rangle =215$ km
s$^{-1}$ and 25 km s$^{-1}$ dispersion (plus some stars at slightly
lower velocities that may also be associated with the
structure). These stars extend the Virgo substructure to more nearby
distances than previously known. The mean metallicity of six of these
bright RR Lyrae is $\langle [$Fe/H$] \rangle =-1.55$, with
$\sigma_{\rm [Fe/H]}=0.15$ dex. Further spectroscopic data of QUEST RR
Lyrae was reported by
\citet{dvz+10}. This study found a VSS grouping with 23 members,
having a mean RV of $\langle V_{\rm GSR} \rangle =129$ km
s$^{-1}$. There are additional stars in a group at higher
($200\lesssim V_{\rm GSR}\lesssim250$ km s$^{-1}$) velocity, but
slightly more nearby (9-15 kpc). These, together with the 10-12 kpc
$\sim200$ km s$^{-1}$ stars from \citet{vjz+08} suggest a velocity
gradient with distance in the VSS region.

\begin{landscape}
\begin{table}[!ht]
\caption{Previous detections of stellar overdensities in Virgo.}
\begin{center}
\begin{tabular}{|l|c|c|c|c|c|l|}

\tableline 
\multicolumn{1}{|c}{source} & \multicolumn{1}{|c}{RA} & \multicolumn{1}{|c}{Dec} & \multicolumn{1}{|c}{distance} & \multicolumn{1}{|c}{$V_{\rm GSR}$} & \multicolumn{1}{|c}{[Fe/H]} & \multicolumn{1}{|c|}{notes} \\
 & \multicolumn{1}{|c}{(degrees)} & \multicolumn{1}{|c}{(degrees)} & \multicolumn{1}{|c}{(kpc)} & \multicolumn{1}{|c}{(km s$^{-1}$)} & \multicolumn{1}{|c|}{(dex)} &  \\
\tableline

Vivas et al. (2001) & [195,200] & [-2.3,0] & $\sim20$ & ... & ... & 5 QUEST RR Lyrae \\
Newberg et al. (2002) & 190.2 & 0.3 & $\sim20$ & ... & ... & SDSS F-turnoff stars \\
Vivas \& Zinn (2003) & 186 & -1 & 19 & ... & ... & QUEST RR Lyrae ("12.4-hour clump") \\
Duffau et al. (2006) & [175,205] & [-2.5,0.0] & [16,20] & 99.8 & -1.86 & spectra of QUEST RR Lyrae ("Virgo Stellar Stream") \\
Vivas \& Zinn (2006) & 189 & -0.8 & 17 ([12,20]) & ... & ... & QUEST RR Lyrae \\
Newberg et al. (2007) & 191.2, 185.6 & -7.8, -0.1 & 18 & 130$\pm10$ & ... & SDSS F-turnoff stars \\
Juric et al. (2008) & [160,210] & [-5,15] & 6-20 & ... & $\lesssim -1.0$ & SDSS "tomography" ("Virgo Overdensity") \\
Keller et al. (2008) & [180,230] & [-12,-2] & [16,19] & ... & ... & SEKBO RR Lyrae \\
Vivas et al. (2008) & [170,210] & [-2.3,0.0] & [8,12] & 215 & -1.55 & spectra of bright ($V<16.1$) QUEST RR Lyrae \\
Keller et al. (2009) & [180,210] & [-15,0] & [17,20] & ... & ... & SEKBO RR Lyrae with SDSS photometry \\
Prior et al. (2009) & [170,210] & [-14,0] & [16,22] & $127\pm10$\tablenotemark{a} & $-1.95\pm0.1$ & spectra of SEKBO RR Lyrae \\
Brink et al. (2010) & 185.9 & -1.0 & ... & 137\tablenotemark{b} & ... & spectroscopy WFC on INT 2.5-meter \\
Duffau et al. (2010) & [180,210] & [-4,0] & [7,21] & 129 & ... & "VSS" group of QUEST RR Lyrae \\
Duffau et al. (2010) & [180,210] & [-4,0] & [7,18] & $\sim220$ & ... & higher-velocity group of QUEST RR Lyrae \\
Keller (2010) & [180,210] & [-2,6] & [12,18] & ... & ... & SDSS subgiants \\
Casey et al. (2012) & [180,194] & [-3,0] & $\sim20$ & $\sim130$ & -2.0 & spectra; Feature B \\
Casey et al. (2012) & [180,194] & [-3,0] & $\sim20$ & $\sim220$ & -1.2 & spectra; Feature C \\

\tableline

\end{tabular}
\end{center}
\tablenotetext{}{Values in brackets represent ranges in which those particular studies found Virgo substructure members.}
\tablenotetext{a}{Note that this study contained some stars at high ($V_{\rm GSR} > 180$ km s$^{-1}$) velocity that were not identified by the authors as VSS members.}
\tablenotetext{b}{We note that the authors' estimate of $V_{\rm GSR} = 137$ km s$^{-1}$ uses only stars with $V_{\rm GSR} < 170$ km~s$^{-1}$; using all stars with $100 < V_{\rm GSR} < 350$ km~s$^{-1}$ yields a median $V_{\rm GSR} = 152$ km s$^{-1}$.}
\label{tab:litdata}
\end{table}
\end{landscape}

The overdensity of RR Lyrae in Virgo has also been studied by the
SEKBO survey, which spans $\sim10\arcdeg$ about the ecliptic.
\citet{kmp+08} show that the Virgo overdensity is visible among SEKBO
RR Lyrae at $R\sim20$ kpc between $180\arcdeg<$RA$<230\arcdeg$.
Photometric and spectroscopic follow-up of SEKBO stars by
\citet{pdk+09} uncovered a handful of additional VSS stars between
16-22 kpc at mean $V_{\rm GSR}=127\pm10$ km s$^{-1}$ and dispersion of
$\sim27$ km s$^{-1}$. These stars have $\langle$[Fe/H]$\rangle =
-1.95\pm0.1$, with scatter of 0.4 dex (note that an additional star at
$V_{\rm GSR}=193$ km s$^{-1}$ and [Fe/H]~=~-1.34 was not included as a
member in their sample). \citet{pdk+09} also calculate a number
density of excess stars and use this to find that in a $\sim760$
deg$^2$ area centered on $(\alpha,\delta)=(186\arcdeg,-4\arcdeg)$ the
total luminosity is $M_V$ = -11.9 (or $M_V = -10.1$ using values more
closely resembling the \citealt{jib+08} distances). The authors claim
that becuase VSS stars have been found between 12-24 kpc with a large
metallicity spread, we must be looking at the remains of a disrupted
dwarf spheroidal galaxy. We also note that this study extended the VSS
to much lower latitudes of $b\sim45\arcdeg$ ($\delta\sim -15\arcdeg$).

An extensive examination of the Virgo substructure was enabled by SDSS
Data Release 5 (DR5) data. \citet{nyc+07} studied Virgo using SDSS
starcounts of BHB stars, blue stragglers, and F-turnoff stars as well
as SEGUE radial velocities of F-type turnoff stars in the region of
the Virgo overdensity. The authors showed using BHB stars and blue
stragglers within 15 kpc of the Sgr orbital plane that Sgr debris in
this region does not pass through the position of S297+63-20.5
identified by \citet{nyr+02}, ruling out the association of the Virgo
structure with leading-arm Sgr debris. Since the SDSS footprint by DR5
covered the entire north Galactic cap (NGC), Newberg et al. were able
to make starcount maps of the NGC using photometrically-selected
F-turnoff stars. These maps showed that the Virgo (S297+63-20.5)
structure is prominent mostly at magnitudes $20<g<21$, spanning at
least $\sim15\arcdeg$ across. \citet{nyc+07} use SEGUE radial
velocities of F-turnoff stars in two plates at
$(l,b)=(300\arcdeg,55\arcdeg)$ and $(288\arcdeg,62\arcdeg)$ to search
for a Virgo RV signature. A clear excess of F stars is seen at
$\langle V_{\rm GSR} \rangle = 130\pm10$ km s$^{-1}$ with stars in the
30 km s$^{-1}$ wide bins adjacent to the peak, suggesting a large
velocity dispersion within the structure.

\citet{bmm10} undertook a spectroscopic survey of 111 faint
($18.0<g_0<21.8$), blue ($-0.1<(g-r)_0<0.65$) turnoff candidates
centered on the VSS, the majority of which seem to be Sagittarius
leading arm stars. However, there is a broad peak at $\langle V_{\rm
GSR} \rangle = 137$ km s$^{-1}$ with dispersion 22 km s$^{-1}$ that is
consistent with the \citet{nyc+07} VSS/VOD detections from F-turnoff
stars. We note also that the mean and dispersion quoted by
\citet{bmm10} was measured using only stars between $105<V_{\rm
GSR}<170$ km s$^{-1}$, but there is a long tail of stars at higher
velocities in their sample.

The final spectroscopic study we mention is that of
\citet{ckd12}. These authors observed a sample of 178 K-giant stars in
four separate fields between $180 < RA < 195\arcdeg$ and $-3 < Dec <
0\arcdeg$. In addition to a clear signal from the Sagittarius tidal
tail at $V_{\rm GSR} < 0$ km s$^{-1}$, they found a significant excess
of high-velocity ($V_{\rm GSR} \gtrsim 70$ km s$^{-1}$) stars. The
authors suggest that these $\sim20$-kpc-distant stars may be two
separate features (denoted ``Feature B'' and ``Feature C'') with $70
\lesssim V_{\rm GSR} \lesssim 250$ km s$^{-1}$.

Another SDSS study by \citet{jib+08} explored the photometric
properties of the overdensity in Virgo. The prominent factor of
$\sim2$ stellar excess spanning over 1000 deg$^2$ in this study was
dubbed the ``Virgo overdensity (VOD)'' by the authors. The VOD extends
from the lower limit of the SDSS footprint at $Z\sim6$ kpc above the
Galactic plane to the outer limit of their study at $Z\sim20$ kpc,
suggesting that it may extend even farther. The feature appears to
narrow at higher $Z$, which suggests that it may extend further into (or
beyond) the plane. Assuming a distance of 10 kpc and 1000 deg$^2$
area, \citet{jib+08} estimated a lower limit on the total luminosity
of $L_r = 0.1\times10^6 L_{\Sun}$, or $M_r = -7.8$. The color-color
distribution of VOD stars suggests that the VOD is more metal-poor
than the thick disk. However, Juric et al. (as well as
\citealt{msw+03} previously) see the VOD in 2MASS-selected M
giants, so there must also be a relatively metal-rich population
associated with the VOD as well. Again, because of its low surface
brightness, large extent, well-defined outline, and low metallicity,
the authors argue that the VOD is the result of a low-metallicity
dwarf galaxy merging with the Milky Way.

Additional work using SDSS-derived photometric metallicity estimates
by \citet{ajb+09} explored the entire NGC region. The VOD in the
region between $270\arcdeg<l<330\arcdeg$ and $60\arcdeg<b<70\arcdeg$
was found by these authors to have median metallicity of
[Fe/H]=-2.0$\pm$0.1. Note, however, that this is for {\it all} dwarf
stars in this region, only some of which are VOD members. An et
al. claim that the fact that the median [Fe/H] derived is the same as
the halo metallicity in the symmetric area of the sky means that the
additional VOD stars are halo-like.

Using SDSS photometrically-selected sub-giants, \citet{k10a} mapped
the north Galactic cap, and resolved the VOD into a number of discrete
overdensities of smaller area than that of \citet{jib+08}. \citet{kdp09}
found a similar result using combined SEKBO RR Lyrae data and SDSS
photometry. The VOD appears to break up into at least three distinct
clumps in that study. This may suggest significant inhomogeneity across
the larger $\sim1000$ deg$^2$ feature of \citet{jib+08}.

A summary of the numerous detections of stellar overdensities in Virgo
discussed in this section is given in Table~\ref{tab:litdata}. From
this table, it is obvious that there is a large complex of structures
in this part of the sky detected using a variety of stellar tracers,
that (based on their similar distances and velocities) may be related
to each other. Each of the surveys presents a data set limited either
in its spatial coverage or based on a modest-sized sample of spectra,
making interpretation difficult.

Understanding the nature of the Virgo overdensities\footnote{Because
the relationship between the various structures including the ``Virgo
Overdensity (VOD)'' and ``Virgo Stellar Stream (VSS)'' in the Virgo
region is unclear, we will simply refer to ``the Virgo stellar
(sub)structure'' throughout this work. We will show later in this
work that all of these structures are likely related, making this
semantic approach reasonable.} and their possible inter-relationship
will ultimately require measurement of the kinematics and/or chemistry
of large numbers of stars in the large area where overdensities have
been detected. In contrast, we focus on a small region of sky, but
where we can take advantage of precise measurements of
three-dimensional kinematics of selected Virgo candidates. This study
is a follow-up to that of
\citeauthor{cgm+09a} (2009; hereafter CD09), which presented the orbit
of a single RR Lyrae consistent with membership in a Virgo
substructure. This star had radial velocity measurements from SDSS and
proper motion measurements from photographic plate data derived from
the survey of Kapteyn's Selected Areas described initially by
\citet{cmg+06}. The 3-D kinematics of this particular star place it
(and, by inference, the substructure of which it is part) on a highly
eccentric orbit, having just passed its pericenter. The authors
considered possible positional associations with Milky Way satellites
beyond Galactic radii of 50 kpc, and found that the unusual globular
cluster NGC 2419 is within 13 kpc of the derived orbit.

The present study revisits the result of CD09 using spectra for
additional stars from the same proper-motion catalog used in the CD09
study. We spectroscopically identify 16 Virgo members, from which we
derive mean proper motions and RVs. From these, we derive an orbit for
the Virgo substructure, and use this orbit to produce $N$-body models
based on the measured kinematics. The paper is structured as follows:
In Section 2 we discuss the data used for this study, which include
proper motions from the Selected Areas survey, SDSS photometry, and
low-resolution spectroscopy obtained on three separate observing
runs. Section 3 outlines the methods we used to select Virgo
substructure members, and compares our data sets to SDSS data from the
surrounding regions of sky. The criteria adopted to identify Virgo
structure members were based on radial velocity, color-magnitude
location, proper motion, and to some extent metallicity, yielding a
final sample of 16 Virgo members. We follow this, in Section 4, with
analysis of the kinematics derived from the identified members. We
derive an orbit based on the mean motion we measure, and use this to
generate an $N$-body simulation that roughly reproduces the (still
poorly-known) characteristics of the Virgo substructure(s). In Section
5, we search for positional and kinematical associations of known
Milky Way satellites and substructures with our derived orbit for
Virgo debris, and discuss a few possible associations. In particular,
our data and subsequent simulations suggest plausible associations of
the Virgo debris with NGC 2419, the Pisces Overdensity, the
\citet{nyg+03} 90-kpc ``Sgr debris'', and a stellar overdensity in the
field of view of the ultrafaint dwarf galaxy SEGUE 1. We conclude by
arguing that we have convincingly shown that all or most of the known
Virgo overdensities are plausibly part of a single disrupted dwarf
galaxy, and that a number of other Milky Way substructures may be
associated with this same progenitor.

\section{The Data}

\subsection{Photometry and Proper Motions}

Proper motions used in this study come from the survey described in
\citet[hereafter CD06]{cmg+06}; we provide a brief introduction here,
and refer the reader to \citetalias{cmg+06} for details. The survey
consists of $\sim$50 fields of $40\arcmin \times 40\arcmin$ spaced
$\sim15$ degrees apart in strips along the celestial equator (i.e.,
$\delta = 0\arcdeg$) and $\delta = \pm15\arcdeg$ (see Fig.~1 in
\citetalias{cmg+06} for a map). These fields are the "Selected Areas
(SAs)" designated by \citet{k06} for his proposed comprehensive study
of the Milky Way. The particular field of view in this study, SA~103,
was also the focus of a previous publication
(\citetalias{cgm+09a}). SA~103 is centered at $(\alpha, \delta)_{2000}
= (178.765\arcdeg, -0.562\arcdeg)$, or $(l, b) = (274.56\arcdeg,
59.17\arcdeg)$. The photographic plates making up our earliest epoch
for proper motions were originally taken between 1909-1912 (analysis
of photometry from these plates appears in \citealt{skv+30}) as part
of the Kapteyn survey at the Mt.~Wilson 60-inch. Matching plates (in
field of view, depth, and plate scale) were taken by S.~Majewski
between 1996-1998 at the Las Campanas du Pont 2.5-meter telescope to
provide a $\sim90$-year baseline for measurement of proper motions.
The Mt.~Wilson 60-inch plates have plate scale $27\farcs12$ mm$^{-1}$,
and the plate scale for du~Pont 2.5-meter data is $10\farcs92$
mm$^{-1}$, with both having a field of view of $40\arcmin \times
40\arcmin$. All of these plates were digitized using the Yale PDS
microdensitometer. The data were supplemented by intermediate-epoch
plates from the Palomar Observatory Sky Survey (POSS-I), taken in the
early 1950s, to extend the limiting magnitude of our proper motions
beyond the depth provided by the Mt.~Wilson plates. Proper motions
were measured by comparing the positions of objects on 1 Mt.~Wilson
plate, 4 POSS-I plates, and 2 du~Pont plates. Each of the Mt.~Wilson
and du~Pont plates contain a long and short exposure offset spatially,
providing 2 measurements per plate. Two separate scans of the POSS-I
plates were analyzed; one from the PMM measuring engine at USNO, and
the other from the PDS microdensitometer at STScI. Thus there is a
total of 14 possible measurements from the 7 plates in SA~103. The
correction from relative to absolute proper motions was done using
four QSOs and 302 background galaxies in the field of view (see
\citetalias{cgm+09a} for details), from which we derived zero points
of $\mu_{\alpha}$ cos $\delta = 3.89\pm0.27$ mas yr$^{-1}$ and
$\mu_{\delta} = 2.31\pm0.30$ mas yr$^{-1}$; these were applied to the
data to place the proper motions in an absolute frame.\footnote{Note
that these differ from the zero points in \citetalias{cgm+09a}, but
are consistent within the errors, and thus make little difference in
our final conclusions. The difference between the measurements is in
the sample of QSOs used in the two, separate measurements.} QSOs are
taken from SDSS QSO catalogs, and galaxies are identified in the plate
scans.  Proper motion uncertainties range from $\sim1$ mas yr$^{-1}$
per star at the bright end (i.e., $g \lesssim 18$) to 2-3 mas
yr$^{-1}$ at the faint ($g \sim 20-21$), limiting magnitude of the
plates.

\begin{figure}[!Ht]
\includegraphics[height=4.5in]{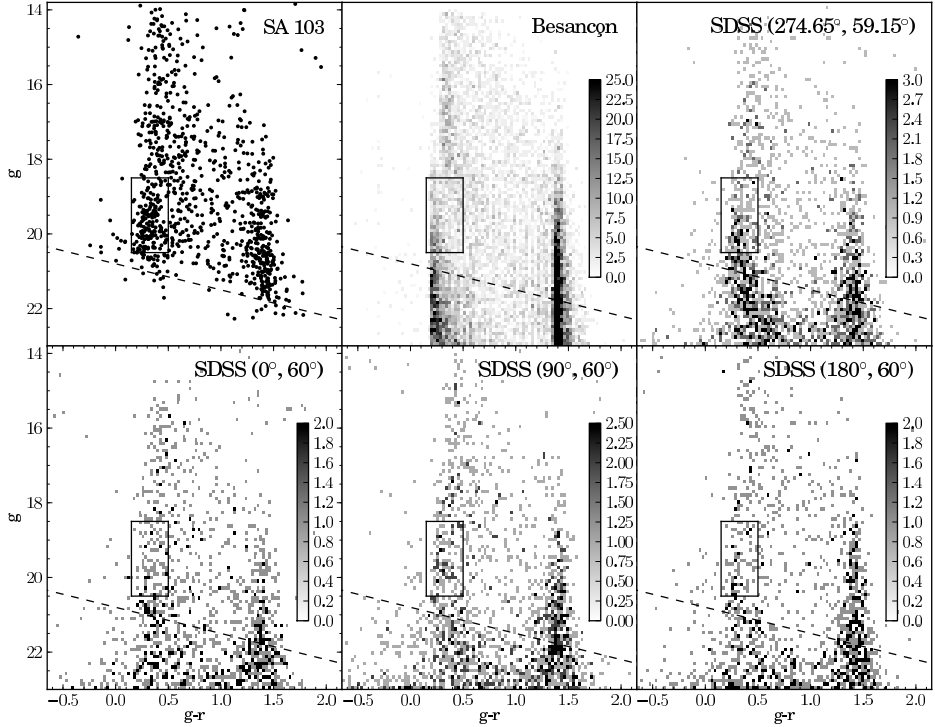}
\caption{Color magnitude diagrams comparing all stars with measured proper motions in our SA~103 data (upper
  left panel) at $(l,b)=(274.6\arcdeg, 59.2\arcdeg)$ with output of
  the Besan\c{c}on model along the same line of sight (upper middle
  panel) and SDSS CMDs from one-degree fields at $b\sim60\arcdeg$ at
  $\sim90\arcdeg$ intervals in Galactic longitude (the upper-right and
  lower three panels, with $l,b$ as labeled; the upper right field is
  SDSS data within the SA~103 field of view). For SA~103 we show each individual star, while all of the other panels are binned Hess diagrams, with the grayscale density values given by the colorbar in each panel. The dashed line in each panel, defined by $g < 0.7(g-r)+20.8$, roughly represents the magnitude limit of the SA~103 data, and is given for reference.
  Assuming the Galactic halo is spherically symmetric at such high
  latitudes, the starcounts of halo stars within the box overplotted
  on each panel should be roughly equal at each longitude.  We find a
  total of 183 stars within the CMD-selection box in the upper-right
  panel (corresponding to SA~103), well above the totals of 93, 130,
  and 85 that are found in each of the fields at $l = 0, 90,
  180\arcdeg$, respectively. When we instead consider the fraction of the total number of stars above the dashed line that are within the boxed region, we find 0.123, 0.125, 0.106, and 0.142 for $l = 0, 90, 180,$ and $274\arcdeg$, respectively. We posit that this excess over the number
  of stars in symmetric fields (all selected from data of equal
  quality and depth) signals the presence of a significant overdensity
  (comprising $\sim30\%$ of the stars in the boxed region) in SA~103.}
\label{fig:CMD}
\end{figure}

Rather than relying on photographic photometry, we took advantage of
the fact that high-quality photometry from SDSS DR7 was available for
the SA~103 area. The photometric data we have used throughout this
work are corrected for reddening and extinction (from \citealt{sfd98},
the mean reddening is E(B-V) = 0.026 along this line of sight).
By comparison of photometric catalogs to the much deeper SDSS data, we
have found that the proper motion catalogs are $86\%$ complete for
point sources at $g = 20.0$, falling to $15\%$ completeness at $g =
21.0$. The limiting magnitude set by the shallow Mt.~Wilson plates
(i.e., for stars with $\sim90$ yr baseline) is $g \sim 20.0$, with
$90\%$ completeness at $g = 17.0$. A color magnitude diagram is given
in the upper left panel of Figure~\ref{fig:CMD} for all stars with
measured proper motions in SA~103. For comparison, the middle panel
shows the predictions of the Besan\c{c}on Galaxy model
(\citealt{rrd+03}) for the same field of view, and the right panel
shows all stars from SDSS DR8 within the same field of view. In the
DR8 panel, an upper main sequence is clearly visible, with a turnoff
at $g\sim19.0$, $g-r \sim 0.3$ -- this likely corresponds to the MSTO
of the Virgo structure identified by \citet{nyr+02}. The same feature
is visible in our SA~103 data, and is not represented in the
Besan\c{c}on model, which bolsters our assertion that this turnoff is
due to a substructure in the Milky Way halo. To further confirm that
this turnoff structure is not expected among Galactic stellar
populations in the SA~103 field of view, we compare SDSS data in
SA~103 to SDSS CMDs at $90\arcdeg$ intervals in longitude and
approximately the same latitude ($b=60\arcdeg$). Assuming the Milky
Way has an axisymmetric stellar halo, these four high-latitude fields
seen in Figure~\ref{fig:CMD} should exhibit the same features among
faint, blue halo stars. Indeed, the three fields at $l=0\arcdeg,
90\arcdeg,$ and $180\arcdeg$ (the bottom panels in
Figure~\ref{fig:CMD}) look similar to each other, but the SA~103
field of view (the upper right panel) shows a clear main sequence
extending downward from $g\sim19, g-r\sim0.3$. Within the overplotted
box at $0.15 < g-r < 0.5$ and $18.5 < g < 20.5$, there are a total of
93, 130, 85, and 183 stars for $l=0, 90, 180, 274\arcdeg$
respectively. These correspond to fractions of 0.123, 0.125, 0.106, and 0.142 of the total number of stars above the dashed line (i.e., with $g < 0.7(g-r) + 20.8$) in each panel of Figure~\ref{fig:CMD}. We thus conclude that the apparent clumping of stars at
$g \gtrsim 18.5$ and $0.15 < g-r < 0.5$ in our SA~103 proper motion
data near $l = 274\arcdeg$ is due to a significant stellar excess that is present along this
line of sight. Based on these starcounts we surmise that $\gtrsim30\%$
of the stars in this CMD-selection region are part of the Virgo
substructure.

\begin{figure}[!ht]
\plotone{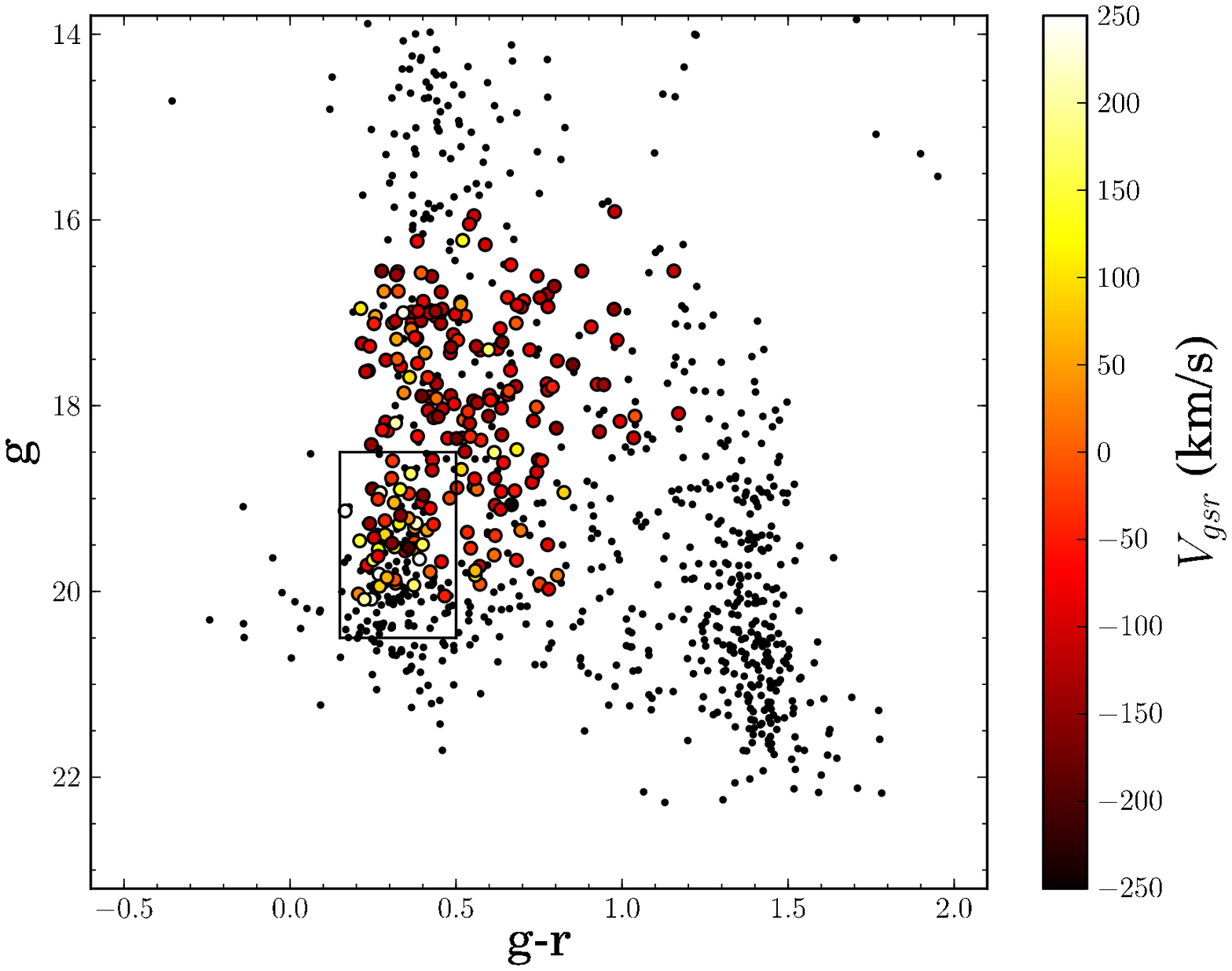}
\caption{Color-magnitude diagram of all stars with measured proper motions in SA~103 shown as small black dots. Stars with spectroscopic data are highlighted by larger circles color-coded by $V_{\rm GSR}$. Brighter colors correspond to higher $V_{\rm GSR}$. The boxed region corresponds to the same Virgo turnoff selection area as in Figure~\ref{fig:CMD}. Note that a larger fraction of the light-colored (i.e., high-velocity) stars is present in the area of the Virgo turnoff than outside this box.}
\label{fig:CMDspectra}
\end{figure}

\subsection{Radial Velocities}

In Figure~\ref{fig:CMD} there is no obvious red giant branch
associated with the MSTO feature, meaning that spectroscopic follow-up
aimed at efficiently identifying and measuring the kinematics of Virgo
substructure members should ideally focus on the high-density turnoff
region. However, many of the spectra we observed for this study were
taken as backup targets during observing time scheduled for other
projects. Thus the spectroscopic targets, shown as large colored
circles in Figure~\ref{fig:CMDspectra}, are not optimally selected and
tend to sample brighter magnitudes. The box overplotted in
Figure~\ref{fig:CMDspectra} ($0.15 < g-r < 0.5$, $18.5 < g < 20.5$;
the same as in Figure~\ref{fig:CMD}) surrounds the apparent MSTO of
the Virgo structure, and thus contains the best spectroscopic
candidates. Of the total of 215 stars for which we obtained spectra,
53 are within this box, with the majority of the remaining spectra at
brighter magnitudes. Many of the stars brighter than $g \sim 18.5$
were selected in the region immediately surrounding the RR Lyrae star
($g = 17.05, g-r = 0.24$) on which the CD09 study focused, with the
hope of identifying additional AGB/HB stars in the Virgo
structure. The remainder of bright targets are redward of the
thin/thick disk MSTO, where Virgo RGB candidates would be expected to
reside.

Spectra of candidates were obtained on three separate observing runs.
The first of these, in June 2009, used the Hydra multifiber
spectrograph on the WIYN 3.5-meter telescope.\footnote{The WIYN
Observatory is a joint facility of the University of
Wisconsin-Madison, Indiana University, Yale University, and the
National Optical Astronomy Observatory.} A single setup consisting of
70 targets was observed at $R\sim1500$ ($3.35$~\AA~at $\lambda =
5200$~\AA~) using the 600-line (``600@10.1'') grating centered at
$\lambda \sim 5700$~\AA~(providing wavelength coverage of
$4300-7100$~\AA). The remaining candidates were observed as backup
targets during runs at Cerro Tololo Inter-American Observatory in
February 2010 and January 2011 using the Hydra multifiber spectrograph
on the 4-meter Blanco telescope. The observing time for these runs was
granted by NOAO to search for QSOs to use as astrometric reference
sources in the field of view of the Carina dwarf spheroidal -- results
from that study will be presented in a future work. We observed at low
resolution ($R\sim850$), using the KPGL2 grating in first order
centered at $\lambda \sim 6000$~\AA.  This setup was chosen to provide
spectra that cover a broad ($\lambda \approx 3400-8200$~\AA)
wavelength range for finding the emission lines for QSOs within a
large redshift window.  Such spectra are not ideal for stellar
kinematics, providing radial velocities of $\sim10-20$ km s$^{-1}$
accuracy (compared to 5-10 km s$^{-1}$ with the WIYN+Hydra setup; both
estimates assume cross-correlations accurate to $\sim\frac{1}{30}$ of
a resolution element). However, in both the 2010 and 2011 runs, our
observations were scheduled in a period when the Carina dSph was
unobservable at the end of the night, so we observed SA~103 to fill
extra time. In February 2010, we observed 3 configurations totaling
188 stars that had spectra of sufficient signal-to-noise ($S/N$ > 10)
to derive velocities.  In January 2011 we observed two additional
SA~103 configurations with the same spectrograph setup, totaling 162
stars with usable spectra. Many stars were targeted on multiple
configurations to allow cross-checking of the radial velocities; thus
the total number of stars with measured velocities in SA~103 is 215.

For all observing, the target configurations were observed multiple
times to enable cosmic-ray removal. The initial pre-processing of 2-D
spectra used the CCDRED package in IRAF.\footnote{IRAF is distributed
  by the National Optical Astronomy Observatory, which is operated by
  the Association of Universities for Research in Astronomy (AURA)
  under cooperative agreement with the National Science Foundation.}
One-dimensional spectra were extracted from the summed 2-D frames
using the DOHYDRA routines (also in IRAF) for multifiber data
reduction. Dispersion solutions were determined using 30-35 emission
lines from arc lamp exposures taken at each Hydra configuration.
Bright radial velocity standards of various spectral types from F
through K (both dwarfs and giants) were observed on each run. Each of
these was observed through multiple fibers, to yield numerous
cross-correlation spectra for RV measurement. Cross-correlation of
these standard spectra against each other was done using the IRAF
FXCOR package; we found that measured velocities of the RV standards
typically agreed with published values to within 1-2 km s$^{-1}$ for
the June 2009 WIYN+Hydra run, and to $\lesssim 4$ km s$^{-1}$ for the
lower-resolution Feb. 2010 and Jan. 2011 CTIO+Hydra observing. The
radial velocities for program stars were then derived by
cross-correlating the object spectra against all of the standard stars
observed on the same run on which the target star was observed.

Radial velocity uncertainties were derived using the \citet{vmo+95}
method, as described in \citet{mcf+06} and \citet{cmc+12}. Because the
Tonry-Davis ratio (TDR; \citealt{td79}) scales with $S/N$, we were
able to use the dependence of the TDR on $S/N$ mapped via multiple
observations of each standard star (which of course have varying
$S/N$) to derive individual RV uncertainties from the measured TDR of
each spectrum. Typical RV uncertainties for individual measurements
are $\sigma_{V} \approx 10-20$ km s$^{-1}$ for spectra having
$S/N\gtrsim$15--20 per Angstrom. For the 100 stars (out of 215 total)
that were observed multiple times, the final radial velocities
represent the error-weighted mean values of individual measurements.
We note that the RV uncertainties based on the TDR method may be
underestimated; comparisons of the velocities of stars observed
multiple times show standard deviations of $25-30$ km s$^{-1}$. As
this includes all RV measurements (including those from low $S/N$
spectra), we estimate that typical uncertainties are $\sim20$ km
s$^{-1}$.

\subsection{Metallicities}

To estimate metallicities from the low-resolution spectra used in this
study, we relied on the well-calibrated Lick spectral indices (see,
e.g., \citealt{wfg+94,f87}). Details of the software pipeline produced
by our group for this purpose will appear in a forthcoming paper
(Carlin et al. 2012, {\it in prep.}). The program derives [Fe/H] based
on fits of the behavior of [Fe/H] as a function of eight Lick Fe
indices and the H$\beta$ index as defined by \citet{wfg+94}. The fits
are calibrated using the known [Fe/H] and spectral indices in the
atlas of \citet[based on the spectra of \citealt{j98}]{s07}. This code
has been shown to yield [Fe/H] measurements of $\sim0.3$ dex precision
(or perhaps better) for low-resolution spectra having $S/N > 20$. For
all stars with multiple measurements of [Fe/H], we adopted the
weighted mean from all measurements. Comparisons of [Fe/H] for the
stars with multiple observations show a standard deviation of
$\sim0.3$ dex in the residuals between individual observations,
suggesting that this is a reasonable value for the typical metallicity
uncertainty.

\section{Virgo Substructure Candidate Selection}
\label{section:candidates}

We have already shown in Section~2.1 that the stellar excess of blue,
faint stars in SA~103 is not due to expected Galactic halo
populations. Our estimate that $\sim30\%$ of the stars within the CMD
box surrounding the turnoff region should be Virgo members suggests
that roughly 15-20 of the 53 spectroscopic targets within that box
should be part of the Virgo substructure, with a handful of additional
RGB and AGB candidates at brighter magnitudes.
Figure~\ref{fig:CMDspectra} shows the CMD for all stars with measured
proper motions in SA~103, highlighting stars having spectra with $S/N
> 10$. Stars are color coded by radial velocity to search for
clumping in $V_{\rm GSR}$. From these data, there appears to be a
higher concentration of high velocity stars (the lighter colored
points) in the boxed region of the CMD. This is consistent with prior
studies claiming that the VSS is comprised of high velocity stars.

\begin{figure}[!Ht]
\plotone{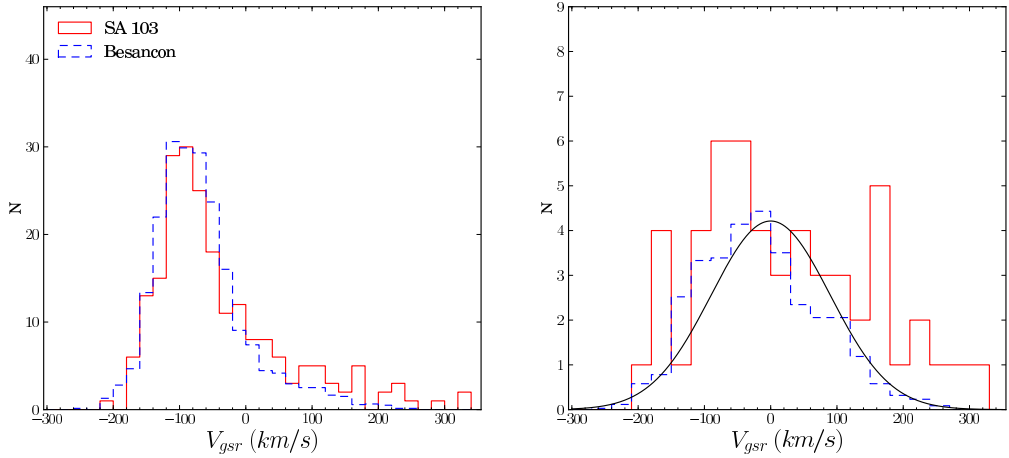}

\caption{The left panel shows the measured $V_{\rm GSR}$ for all stars
  with spectroscopic data in SA~103 as the solid (red) histogram. The
  right panel contains only stars within the CMD box outlined in
  Figures~\ref{fig:CMD} and \ref{fig:CMDspectra}.  The blue dashed
  line in both panels represents the velocities predicted by the
  Besan\c{c}on model in the field of SA~103 within the same color and magnitude ranges as the SA~103 stars. On the left, the
  Besan\c{c}on distribution has been scaled such that the area under
  the histogram within the region between $-300 < V_{\rm GSR} < 100$
  km~s$^{-1}$ equals that of our SA~103 sample. On the right, the
  Besan\c{c}on model is scaled by the fraction of the total number of
  stars within the CMD-selection box among SDSS DR8 data in this field
  of view that were observed; i.e., since we obtained spectra of 53 of
  183 total stars, the Besan\c{c}on histogram was scaled down by a
  factor of 53/183 to match the number of SA~103 stars observed
  spectroscopically. In the left panel, we include only stars between
  $16 < g < 20, 0.2 < g-r < 0.9$ from the Besan\c{c}on model to
  roughly mimic the color and magnitude range covered by the SA~103
  Hydra spectra. On the right, a Gaussian centered at $V_{\rm GSR} =
  0$ km~s$^{-1}$ with a $\sigma = 90$ km~s$^{-1}$ (representing the
  typical Galactic halo distribution, with $\sigma=90$ km~s$^{-1}$
  derived from a fit to the stars identified as halo members in the
  Besan\c{c}on model) is also superposed, again normalized to the
  area of the SA~103 histogram within the region mentioned above. In
  the right panel (i.e., within the main sequence turnoff region
  contained by the CMD selection box) there is a clear excess of
  SA~103 stars over both the model and the halo predictions at $V_{\rm
  GSR} \gtrsim 120$ km~s$^{-1}$. This excess is not confined to a
  narrow, kinematically-cold peak, but rather spreads over a large
  range in velocity. Note also that there may be stellar excesses
  present at $-100 \lesssim V_{\rm GSR} \lesssim -50$ km s$^{-1}$ and
  $V_{\rm GSR} \sim -170$ km s$^{-1}$; these correspond to similar
  excesses seen in this region of the sky by \citet{nyc+07} and
  \citet{bmm10} and suggested to be related to the Sagittarius tidal
  stream. }

\label{fig:SA103histvgsr}
\end{figure}

Figure~\ref{fig:SA103histvgsr} shows histograms of the line-of-sight
$V_{\rm GSR}$\footnote{Velocity in the Galactic Standard of Rest,
defined as $V_{\rm GSR} = V_{\rm helio} +
9.0\cos{b}\cos{l}+232.0\cos{b}\sin{l}+7.0\sin{b}$.} of all stars in
the SA~103 field; the left panel depicts all stars that have
spectroscopic data with $S/N > 10$ (the color-coded points highlighted
in Figure~\ref{fig:CMDspectra}). In the right panel we show only a
subset of faint, blue stars corresponding to the boxed turnoff region
in Figures~\ref{fig:CMD} and \ref{fig:CMDspectra}.  In the absence of
substructure, stars within this box will be primarily Galactic halo
turnoff stars. These halo stars should have a velocity distribution
centered at $V_{\rm GSR} \sim 0$ km s$^{-1}$. A Gaussian fit to all
halo stars from the Besan\c{c}on model for the SA~103 field yields a
velocity dispersion for the halo of $\sim90$ km s$^{-1}$, which we
overplot in the right panel as a reference. The histograms in each
panel are compared to the aforementioned Besan\c{c}on model (the
dashed blue lines) corresponding to the same field. In the left panel,
the model was normalized so that the number of stars in the velocity
range $-300 < V_{\rm GSR} < 100$ km s$^{-1}$ is equal to the number of
stars in the same range in SA~103. This velocity range was chosen so
that the normalization includes only the region of the SA~103 data
that is free from apparent substructure. In the right panel, the
Besan\c{c}on model counts were scaled by the fraction of the total
number of stars in SA~103 within the CMD-selection box that were
observed; i.e., because we obtained spectra of 53 of 183 total stars
in the turnoff box (determined from SDSS DR8 data), the Besan\c{c}on
histogram was scaled down by a factor of 53/183. If the SA~103 data
contained only halo stars, the Besan\c{c}on model scaled in this
manner should fairly accurately predict the total number of stars in
the right panel of Figure~\ref{fig:SA103histvgsr}; clearly there are more
stars in SA~103 than predicted by the model. There seems to be an
excess of stars in the region of $V_{\rm GSR} \gtrsim 100$ km s$^{-1}$
relative to both the Besan\c{c}on model and the theoretical halo
distribution. This excess appears to have a narrow peak centered at a
value somewhat greater than the $\langle V_{\rm GSR}\rangle = 137$ km
s$^{-1}$ found by \citet{bmm10}, though there also appear to be extra
stars over a much broader range of high velocities. Note also that the
peaks at $V_{\rm GSR} \sim -75$ km s$^{-1}$ and $\sim -170$ km
s$^{-1}$ seen by \citet{nyc+07}, \citet{bmm10}, and \citet{ckd12} are
apparent in this field of view; these peaks have been suggested to
contain Sagittarius tidal debris.
 
\begin{figure}[!Ht]
\begin{center}
\includegraphics[height=3.0in]{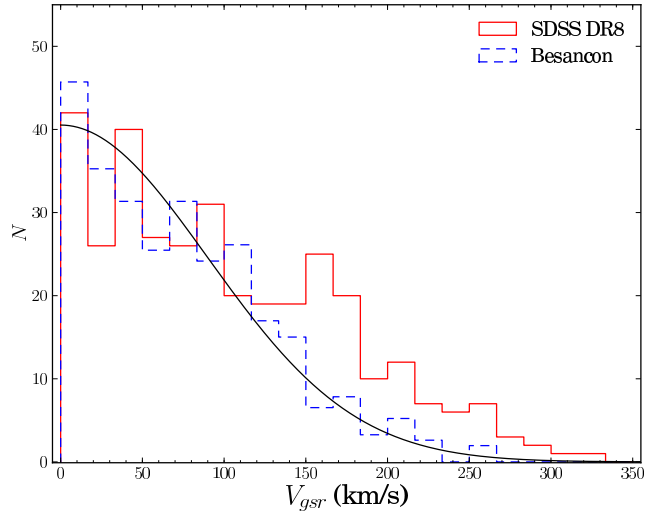}
\caption{Galactic rest-frame radial velocities of stars selected from $265\arcdeg<l<283\arcdeg$ and $50\arcdeg<b<68\arcdeg$ from SDSS DR8 within the same CMD box (i.e, between $0.15<g-r<0.5$ and $18.5<g<20.5$) as in SA~103.  The Besan\c{c}on model from the 1 deg$^2$ field of SA~103 is shown as a dashed blue line for comparison. Here, the Besan\c{c}on model was normalized so that the areas of the histograms are equal within the region of $0 < V_{\rm GSR} < 100$ km s$^{-1}$ (the data was cut off at $V_{\rm GSR} > 0$ km s$^{-1}$ because many Sagittarius stream stars appear in this region of sky at negative velocities). A Gaussian centered at 0 km s$^{-1}$ with a dispersion of $90$ km s$^{-1}$ (representing the Milky Way halo) is plotted as a solid black line. There is a distinct excess of stars in every bin at $V_{\rm GSR} \gtrsim 120$ km s$^{-1}$ relative to both the Besan\c{c}on predictions and the halo Gaussian.}
\label{fig:DR8histvgsr}
\end{center}
\end{figure}

To verify the existence of the apparent high-velocity excess, a
comparison sample of all stars between $0.15<g-r<0.5$ and
$18.5<g<20.5$ that were observed spectroscopically by SEGUE from
$265\arcdeg<l<283\arcdeg$ and $50\arcdeg<b<68\arcdeg$ (i.e., a $>300$
deg$^2$ area surrounding SA~103) was extracted from SDSS DR8. A
histogram of velocities for these stars is seen in
Figure~\ref{fig:DR8histvgsr}. Because of the prominence of the
Sagittarius stream at negative $V_{\rm GSR}$ in this region, we show
only $V_{\rm GSR} > 0$ km s$^{-1}$ in the figure. For comparison, we
overlay (as a dashed, blue line) the Besan\c{c}on distribution from
the smaller SA~103 field, scaled so that the number of stars between
$0 < V_{\rm GSR} < 100$ km s$^{-1}$ is equal to that in the SDSS
histogram. There may be differences in the model between the smaller
SA~103 field and the larger SDSS region, but such differences should
be small because the expanded field is centered in SA~103. To
represent expected Galactic halo populations, we also overplot a
Gaussian centered at $0$ km s$^{-1}$ with a $\sigma=90$ km
s$^{-1}$. As in our SA~103 data, the DR8 stars exhibit a clear excess
in each histogram bin at high ($V_{\rm GSR} \gtrsim 120$ km s$^{-1}$)
velocities relative to the model and the predicted halo
distribution. Because the SEGUE selection function has a patchy
distribution in color and magnitude, these stars are likely not a
complete representation of the velocity distribution in this
region. The excess of stars at high $V_{\rm GSR}$ is present even in
this sample, leading us to conclude that (a) the apparent excess in
SA~103 is due to a real overdensity at high ($V_{\rm GSR}
\gtrsim 120$ km s$^{-1}$) velocities, and (b) the large spread of
velocity in the excess is intrinsic, and not an artifact of the low
resolution of our Hydra data.

Once a rough idea of the velocity of Virgo candidate stars was known,
we proceeded with a detailed candidate selection. The steps we
undertook to identify a sample of Virgo substructure members were:

\begin{enumerate}

\item Select all stars with $40 < V_{\rm GSR} < 350$ km s$^{-1}$. This
  broad range in velocity was chosen to include all Virgo substructure members, but
  will also contain some MW contaminants. A total of 38 stars were
  selected in this manner.

\item Narrow this sample to include only
  stars with at least 3 position measurements spanning at least a
  40-yr time baseline to ensure that their proper motions are
  well-measured. In Figure~\ref{fig:VPDvcut}, which compares proper
  motions of sample 1 in the left panel to those of the Besan\c{c}on
  model on the right, there is a clear clumping of the proper motions
  about the median value of $(\mu_\alpha$ cos $\delta, \mu_\delta) =
  (-4.8, -0.2)$ mas yr$^{-1}$. Furthermore, this clump is clearly
  offset from the expected PMs of halo stars predicted by the
  Besan\c{c}on model. We thus choose to keep only those stars with
  $\Delta\mu < 8$ mas yr$^{-1}$, where $\Delta\mu$ is the offset from
  the median PMs of the stars in selection 1. This selection retains
  30 stars.

\begin{figure}[!Ht]
\plotone{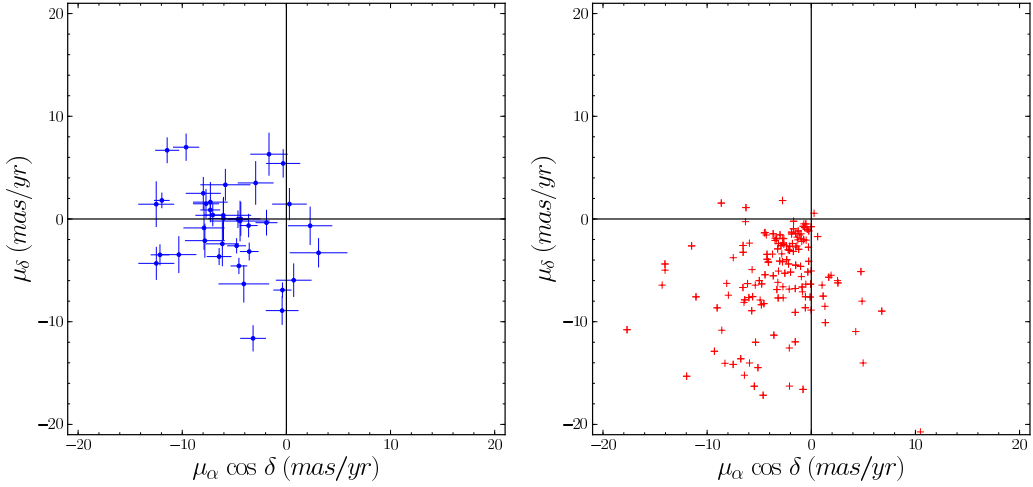}
\caption{The left panel shows the proper motion vector point diagram (VPD) of all stars with
  well-measured PMs and $40<V_{\rm GSR}<350$ km s$^{-1}$ in SA~103. On
  the right we show stars from five realizations of the Besan\c{c}on
  model within the same range of $V_{\rm GSR}$ as the left
  panel. Color and magnitude cuts from $16 < g_0 < 20$ and $0.25 <
  (g-r)_0 < 0.7$ were applied to the Besan\c{c}on points since
  spectroscopic data from SA~103 are roughly limited to within this
  region. The SA~103 stars predominantly occupy a much different
  region than the predicted Galactic halo stars in this field. The
  mean proper motion of the stars in the velocity-selected SA~103
  sample is $(\mu_{\alpha} \cos \delta, \mu_\delta)\sim(-5.2, 0.0)$
  mas yr$^{-1}$.}
\label{fig:VPDvcut}
\end{figure}

\item Calculate the offset of each star in sample 2 from an
  appropriate color-magnitude sequence for the sample 2 stars. For
  this purpose we chose to use the empirical SDSS globular cluster
  ridgelines from \citet{ajc+08}. This mitigates any uncertainties
  that would arise from the use of theoretical isochrones (see
  \citealt{ajc+08} for illustration of the differences between
  empirical and theoretical ridgelines for MW clusters). Implicit in
  our use of globular cluster ridgelines is the assumption that the
  Virgo stars come from a predominantly old ($>10$ Gyr) stellar
  population; this is a reasonable assumption if the stars came from a
  dSph galaxy. The median metallicity of the 23 stars from sample 2
  with spectra having $S/N > 15$ is [Fe/H] $\approx$ -1.1 with
  $\sigma_{\rm [Fe/H]} = 0.6$. We used the SDSS ($g$, $g-r$) ridgeline
  from \citet{ajc+08} for the globular cluster with metallicity
  nearest this value -- that of M5, a cluster with [Fe/H] = -1.27. We
  plotted the dereddened CMD of all stars from SDSS DR8 in the SA~103
  field of view (the stars that make up the upper-right panel of
  Figure~\ref{fig:CMD}), highlighting the candidates in sample 2. We
  then overplotted the M5 ridgeline, shifting it in distance until we
  visually determined a ``best-fit'' distance of $\sim14$ kpc.
  We excised all stars more than 0.2 magnitudes from the 14-kpc M5
  locus in ($g, g-r$) at magnitudes fainter than $g = 17.5$; this
  selection criterion serves to remove stars well off the expected CMD locus
  based on the observed turnoff, and the magnitude limit
  means that this selection is only applied for the turnoff, subgiant
  branch, and lower RGB. The upper RGB and AGB are not included
  because these regions are much more sensitive to differences in the
  (unknown quantities) age and [Fe/H]. Although our
  derived metallicity is higher than most other estimates for
  Virgo-substructure stars, and the distance we have found is slightly
  nearer than most findings, this has little effect on the CMD
  filtering. A generous width of $\pm$0.2 magnitudes from the
  ridgeline would likely encompass all of the same stars if we were to
  change the parameters of the ridgeline used for filtering. This
  additional cut reduces the sample to 25 stars.

\item After selection number 3, the proper motions of most of the
  remaining stars were clumped tightly about $(\mu_\alpha$ cos
  $\delta, \mu_\delta) = (-5.1, -0.2)$, with some obvious outliers at
  the edges of the distribution. This is illustrated in
  Figure~\ref{fig:VPDselection}, which shows the proper motion vector
  point diagrams (VPDs) of all stars with $40 < V_{\rm GSR} < 350$ km
  s$^{-1}$ (i.e., sample 1) in the left panel, color-coded by their
  measured radial velocities. We noted that many of these stars in the
  outer regions of the PM distribution were also at the extreme ends
  of the $V_{\rm GSR}$ selection range, suggesting that they were
  Galactic interlopers, or had suspect proper motion measurements
  (typically having only 4-5 of the 14 possible position
  measurements). We thus chose to remove all stars with $\Delta\mu <
  5$ mas yr$^{-1}$ from sample 3, where the 5 mas yr$^{-1}$ limit was
  chosen by examination of the left panel of
  Figure~\ref{fig:VPDselection}. The VPD of the remaining sample of 16
  stars is shown in the right panel of Figure~\ref{fig:VPDselection}.
  None of the weighted mean proper motions and radial velocities of
  samples 3 and 4 varied more than $1\sigma$ from the sample 2
  measurements, suggesting that the somewhat arbitrary choices of
  exclusion criteria were not subjectively biasing our selections
  strongly. This final sample of 16 stars is highlighted in the CMD in
  Figure~\ref{fig:CMDselection}. The CMD shows all stars in sample 3
  as filled symbols, color-coded by radial velocity. The 16 Virgo
  candidates constituting sample 4 are filled circles, and stars
  excluded by the proper motion cut are triangles. The RR Lyrae star
  RR 167, a known VSS member from \citet{cgm+09a}, is the cyan-colored
  circle at $g\sim17, g-r\sim0.2$, and the two stars with $V_{\rm GSR}
  > 300$ km s$^{-1}$ are shown as green filled hexagons. The overlaid
  isochrones from \citet{ggo+04} are for 10 Gyr populations, with
  [Fe/H]=-1.2, and distances of 14, 17, and 20 kpc (spanning the range
  of most Virgo substructure detections from the literature). The 14
  kpc isochrone appears to be the best match, and the 20 kpc ridgeline
  is clearly discrepant with the Virgo candidates.

\end{enumerate}

\begin{figure}[!Ht]
\plotone{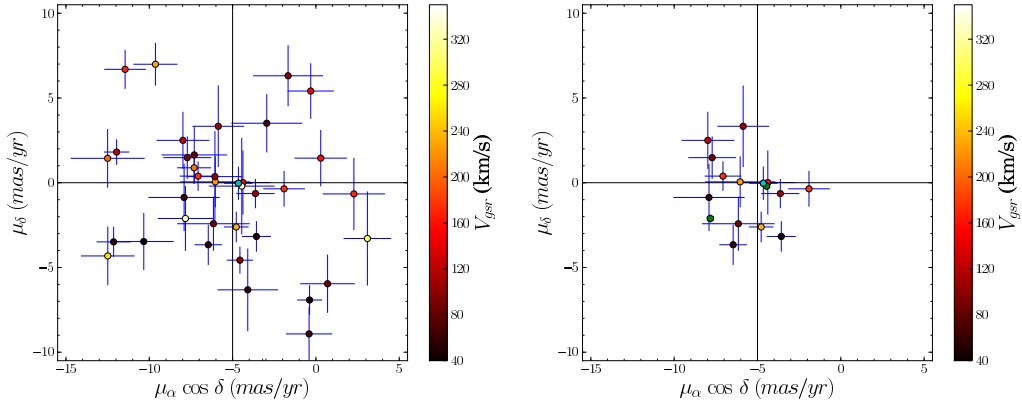}
\caption{The panels show proper motions of Virgo substructure candidates color coded by $V_{\rm GSR}$.  The panels on the left show stars with $40 < V_{\rm GSR} < 350$ km s$^{-1}$, while the ones on the right show just the final selection.  Again, RR 167 is highlighted in cyan.}
\label{fig:VPDselection}
\end{figure}
	
\begin{figure}[!Ht]
\plotone{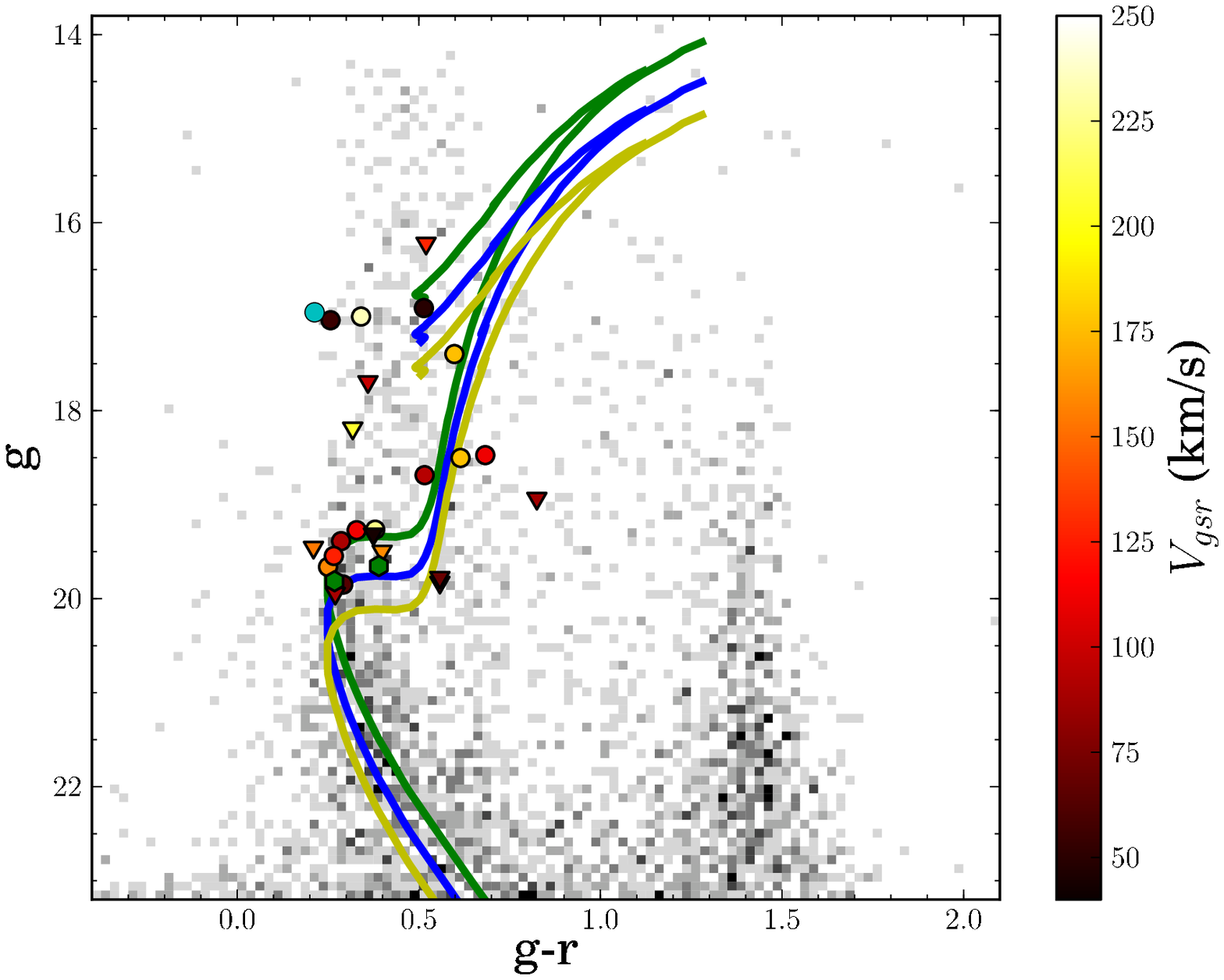}

\caption{Dereddened color magnitude diagram of all stars with $40 < V_{\rm GSR} < 350$ km~s$^{-1}$, $-12 < \mu_{\alpha}\cos\delta < 2$ mas~yr$^{-1}$, $-6 < \mu_{\delta} < 6$ mas~yr$^{-1}$, and measured on at least 3 photographic plates highlighted as circles and triangles. Symbols are color coded by the measured $V_{\rm GSR}$. The greyscale Hess diagram in the background represents stars detected within the field using SDSS DR8 data. Isochrones of age 10 Gyr, metallicity [Fe/H] = -1.2, and distances 14, 17, and 20 kpc from \citet{ggo+04} are superposed in green, blue, and yellow, respectively. Stars that were removed from the sample based on CMD position and proper motion are shown as triangles.  The stars in the final selection are shown as filled circles. RR 167, an RR Lyrae star shown to be a VSS member \citep{cgm+09a}, is highlighted in cyan, and the two stars at high ($V_{\rm GSR} > 300$ km~s$^{-1}$) velocities are filled green symbols.}

\label{fig:CMDselection}
\end{figure}

We note that the final sample of 16 stars contains three stars at
$g\sim17, (g-r) < 0.4$ that, if they are members, must be horizontal
branch stars (specifically, RR Lyrae). One of these three stars, known
as RR 167, has already been shown by \citet{cgm+09a} to be a member of
the VSS, so we believe that it is reasonable to include the other two
in our sample. (Note that although these stars may not have been
detected as RR Lyrae in the QUEST catalog of \citet{vza+04}, the
$\sim70\%$ completeness of QUEST for RRc-type stars in this region of
sky suggests that some bona fide RR Lyrae may have been missed by
QUEST.)  There are also two stars in the sample at rather high RVs of
$V_{\rm GSR} > 300$ km s$^{-1}$, shown as filled green hexagons at
$g=19.75$ and $g=19.91$; these two stars have proper motions
consistent with membership, and because there is not a narrow peak in
the RVs due to Virgo stars, we choose to retain them in our
samples. However, these two high-velocity stars appear to
lie below the MSTO of the 14-kpc isochrone, and thus may be more
distant than the majority of the Virgo members. This may signify
extension of the Virgo structure along the line of sight, including
possibly a velocity gradient with distance.

\begin{figure}[!Ht]
\plotone{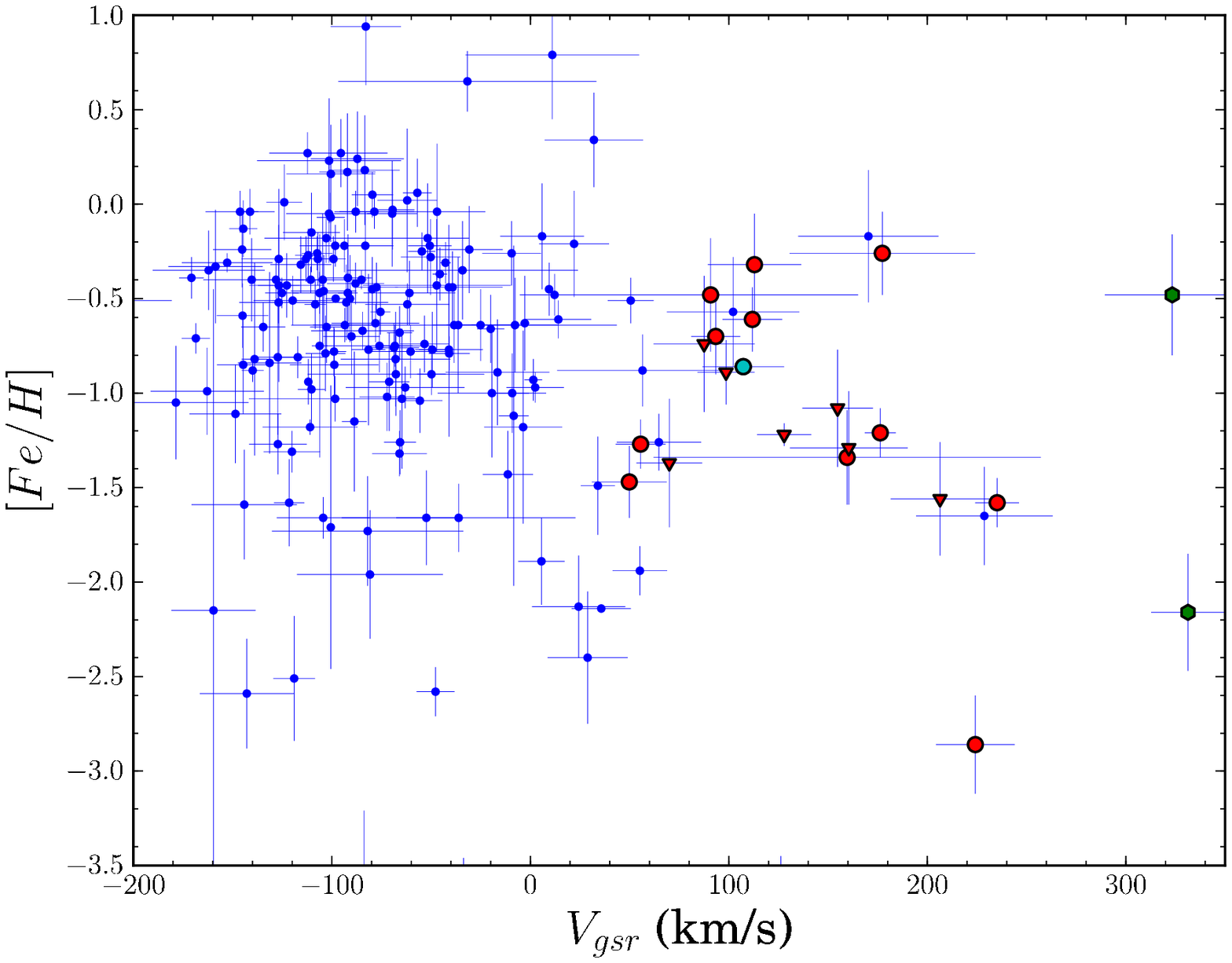}

\caption{Metallicity ([Fe/H]) vs. $V_{\rm GSR}$ for all stars in our
  sample having spectra with $S/N > 15$. The majority of the stars
  between $-150 < V_{\rm GSR} < 50$ km s$^{-1}$ are likely MW
  thin/thick-disk stars with [Fe/H]$ > -1.0$. The large filled circles
  and triangles are the stars in our initial Virgo-candidate velocity
  selection. Triangles are stars that were removed from the sample by
  subsequent selection criteria, leaving only the large filled circles
  as the final Virgo candidates. The two points at the far right side
  have been colored green to distinguish these $V_{\rm GSR} > 300$ km
  s$^{-1}$ stars from the rest of the sample (as in
  Figure~\ref{fig:CMDspectra}). The mean metallicity of the final
  sample is [Fe/H] = -1.1.}
  
\label{fig:fehvgsr}
\end{figure} 

Of the 16 stars in the ``final'' sample, 11 had spectra with $S/N >
20$, and all 16 had $S/N > 14$. From these, we derived metallicities
with $\sim$0.3-dex precision; Figure~\ref{fig:fehvgsr} shows
metallicities for all SA~103 stars with spectroscopic $S/N > 15$ as a
function of velocity. The first selection of $40 < V_{\rm GSR} < 350$
km s$^{-1}$ Virgo candidates are highlighted with larger (red/green)
symbols; large red circles are the final Virgo sample, the two green
circles are the stars flagged in Figure~\ref{fig:CMDspectra} for
having $V_{\rm GSR} > 300$ km s$^{-1}$, and the downward-facing
triangles are stars removed from the final sample based on proper
motion and CMD criteria.  One star from the final sample does not
appear in the figure because of what is likely a spurious
[Fe/H]=-3.75. The 15 stars at [Fe/H]$ > -3.5$ yield a mean value of
$\langle$[Fe/H]$\rangle = -1.1 \pm 0.2$, with $\sigma_{[Fe/H]} = 0.7$
(note that this becomes $\langle$[Fe/H]$\rangle = -1.0 \pm 0.1$,
$\sigma_{[Fe/H]} = 0.6$ if the single star at [Fe/H]$\sim$-2.8 is
excluded).  This metallicity for the Virgo candidates is consistent with
that found from isochrone fitting, and more metal-poor than the
predominantly thick-disk stars toward the left side (i.e., with $-150
< V_{\rm GSR} < 50$ km s$^{-1}$) of Figure~\ref{fig:fehvgsr}.

\section{Kinematics of Virgo Debris}

Statistical kinematical information for the 16 identified Virgo substructure
candidates is given in Table~\ref{tab:stats}. Shown in the table are
the mean spatial coordinates and error-weighted mean proper motions,
radial velocities, and metallicities.  The best fit kinematical
parameters were converted from proper motion and radial velocity to
right-handed Galactocentric Cartesian velocities using the method
described in \citet{js87}.  A Sun-Galactic center distance of 8.0 kpc
was used, as well as solar velocity components in the Galactic rest
frame of $(U, V, W) = (9, 232, 7)$ km s$^{-1}$. The resulting Galactic
Cartesian velocity components of the Virgo debris are $(U, V, W) =
(-264.1, -331.1, 123.9)\pm(65.7, 50.5, 33.0)$ km s$^{-1}$.

These best fit kinematic parameters were used to generate an orbital
model.  The Galactic gravitational potential adopted is that from
\citet{wnz+09} (which is, in turn, based on \citealt{ljm05}): a three
component potential that consists of a \citet{mn75} disk, a Hernquist
spheroid representing the bulge, and a logarithmic halo. The Galactic
potential parameters are the same as those in \citet{ljm05}. Orbits
were constructed using the mkorbit tool of the NEMO Stellar Dynamics
Toolbox (\citealt{t95}). The orbit was integrated for 3 Gyr both
forward and backward using the values tabulated.

\begin{table}
\begin{center}
\caption{Adopted Positions and Weighted Mean Kinematics of 16 Virgo Substructure Candidates \label{tab:stats}}
\begin{tabular}{lr}
\\
\tableline
\\
\multicolumn{1}{l}{Parameter} & \multicolumn{1}{r}{Value} \\
\tableline

\\
RA ($\arcdeg$) & 178.749 \\
Dec ($\arcdeg$) & -0.586  \\
l ($\arcdeg$) & 274.557  \\
b ($\arcdeg$) & 59.118  \\
distance (kpc) & $14\pm3$ \\
$\mu_{\alpha}\cos\delta$ (mas yr$^{-1}$) & $-5.24\pm0.43$  \\ 
$\mu_{\delta}$ (mas yr$^{-1}$) & $-0.91\pm0.46$  \\
$V_{\rm GSR}$ (km s$^{-1}$) & $152.6\pm22.0$\tablenotemark{a}  \\
$[$Fe/H$]$ & -1.1  \\
\tableline
\end{tabular}
\end{center}

\tablenotetext{a}{The $V_{\rm GSR}$ value quoted here is the weighted mean of 13 stars, after excluding the 3 likely RR Lyrae at $g\sim17$. However, we note that if these candidate RR Lyrae stars have any large-amplitude RV variability, it seems to have no impact on our mean kinematics -- including all 16 stars yields $V_{\rm GSR} = 151.8\pm20.4$ km s$^{-1}$.}
\end{table}

\subsection{Orbit}
\label{subsection:orbit}

\begin{figure}[!Ht]
\plotone{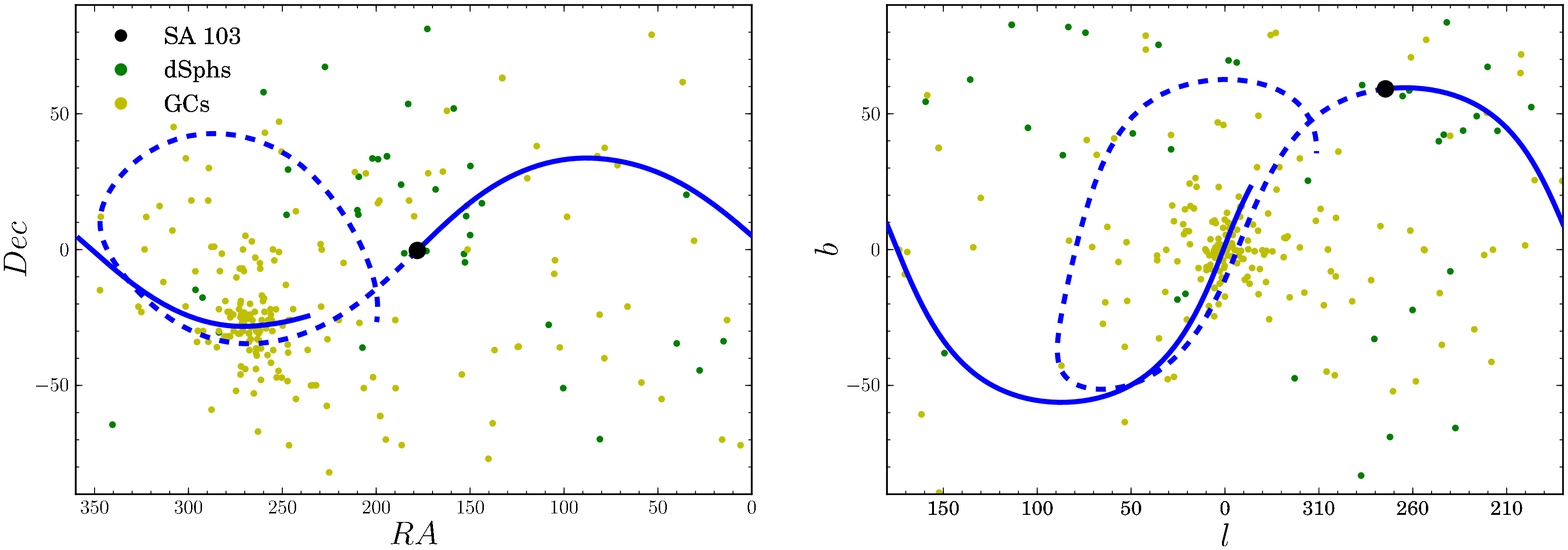}

\caption{The orbit resulting from our measured Virgo kinematics,
  integrated for 1~Gyr forward and backward, is shown as the thick
  blue line in each of the panels above. The dashed portion is the
  backward integration (i.e., the path of the infalling progenitor),
  and the solid part represents the forward integration of the
  orbit. The left panel depicts the orbital path in equatorial
  coordinates, and the right panel in Galactic coordinates. The
  kinematics used to generate this orbit are given in
  Table~\ref{tab:stats}. The location of SA~103 is shown as a large
  black dot at $(\alpha, \delta) = (178.7, -0.6)\arcdeg$, or $(l, b) =
  (274.6, 59.1)\arcdeg$. The positions of known globular clusters and
  dwarf galaxies are plotted in the background as yellow and green
  points respectively.}

\label{fig:lbplot}
\end{figure}

\begin{figure}[!Ht]
\plotone{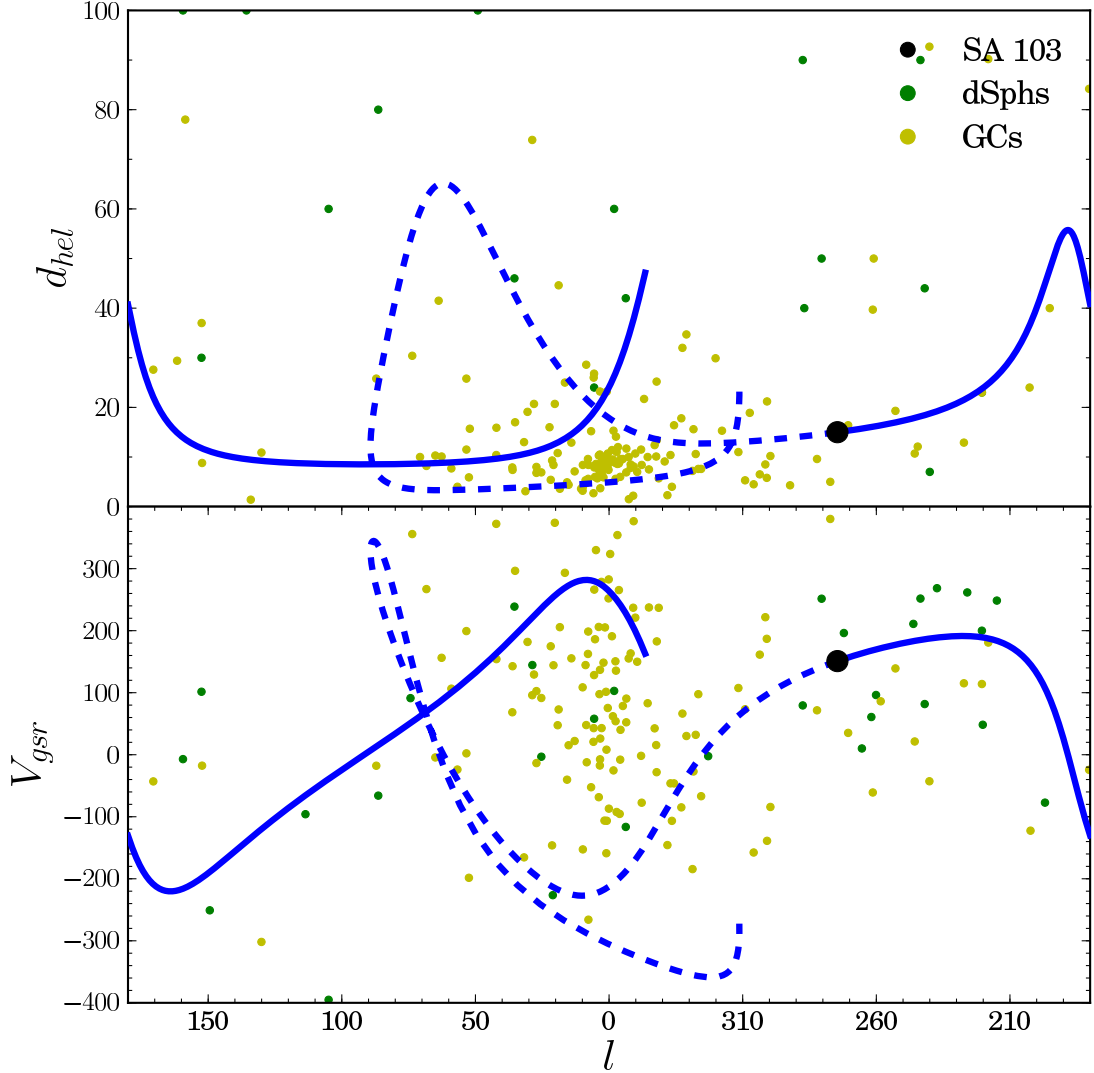}
\caption{Derived orbit as a function of distance and $V_{\rm GSR}$.
  The upper panel shows heliocentric distance vs. Galactic longitude
  for the derived Virgo substructure orbit (thick blue line). In the
  lower panel, we give $V_{\rm GSR}$ vs. Galactic longitude. Plots are
  labeled similarly as in Figure~\ref{fig:lbplot}, with Galactic
  satellites (dSphs and GCs) shown as green/yellow dots. It is evident
  that the current Virgo debris is near orbital perigalacticon and has
  just passed fairly near the Galactic center.}

\label{fig:dvplot}
\end{figure}

Figure~\ref{fig:lbplot} shows the resulting orbit in both equatorial
and Galactic coordinates.  Milky Way dwarf spheroidal galaxies (dSphs)
and globular clusters (GCs)\footnote{From the compilation of \citet{dn09}, available at \url{http://homepages.rpi.edu/~newbeh/mwstructure/MilkyWaySpheroidSubstructure.html}.} are overplotted for later reference with
respect to finding possible orbital associations and/or the
progenitor. The orbit has apo- and peri-centers of $R_{\rm a} =
52^{+87}_{-24}$~kpc and $R_{\rm p} = 5.6^{+0.9}_{-0.8}$~kpc, with
eccentricity $e \equiv (R_{\rm a}-R_{\rm p})/(R_{\rm a}+R_{\rm p}) =
0.81^{+0.10}_{-0.27}$ (errors on these quantities are derived by
propagating the uncertainties in the kinematics). The orbital
inclination with respect to the Galactic plane is $i \equiv \arcsin
(\vert z_{\rm max} \vert / R_{\rm zmax}) = 54\pm3\arcdeg$, where
$\vert z_{\rm max} \vert$ is the maximum height above or below the
plane, and $R_{\rm zmax}$ is the distance from the Galactic center
where the orbit passes through $\vert z_{\rm max} \vert$. The orbital
period (measured between successive apocenters) is $P_{\rm orb} =
0.7^{+1.2}_{-0.3}$~Gyr.  Plots of the orbit in distance and $V_{\rm
GSR}$ vs. Galactic longitude are shown in Figure~\ref{fig:dvplot}.
Again the dwarf galaxies and globular clusters are plotted in the
background for reference. The high eccentricity ($e \sim 0.8$) of this
orbit is evident, with the satellite's trajectory having just passed
pericenter after swooping in from the outer halo.

\begin{figure}[!Ht]
\begin{center}
\includegraphics[height=5.5in,trim=0.2in 0.2in 0.2in 0.2in,clip]{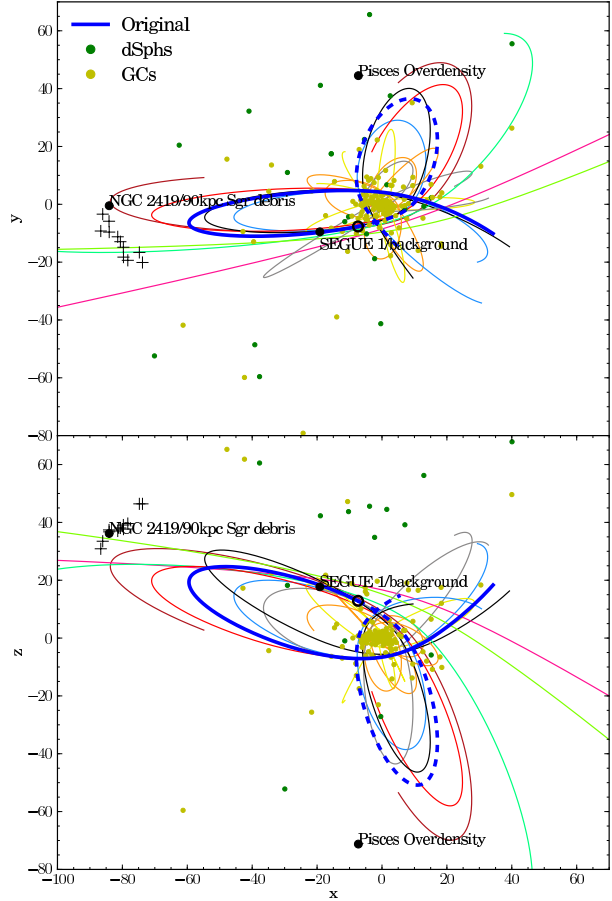}

\caption{The upper panel shows the projection of the Virgo structure orbit onto the
  Galactic $XY$ plane, and the lower panel gives the $XZ$ projection.
  Thick blue lines are the orbit derived from our mean kinematics;
  solid blue lines are the forward integration, and dashed blue lines
  the backward integration. The thinner lines of other colors
  represent orbits generated by selecting from a Gaussian $1-\sigma$
  distribution of the errors in the three velocity components and the
  distance, and illustrate the constraints placed on the Virgo
  progenitor orbit by our measurements. The Sun is at
  $(X,Y,Z)=(-8,0,0)$ kpc in this coordinate system, with positive $X$
  pointing in the direction from the Sun to the Galactic center. The
  position of SA~103 is represented as a large black circle. We also
  highlight as a large black dot the globular cluster NGC~2419, which
  was tentatively associated with the VSS by \citetalias{cgm+09a},
  along with the ``90 kpc Sgr debris'' from \citet{nyg+03} as plus
  signs. Two other substructures found to lie near our derived orbit
  -- SEGUE 1 (or, more specifically, debris in the field of view of
  this ultra-faint dSph) and the Pisces Overdensity -- are also given
  as large black dots.}

\label{fig:xy}
\end{center}
\end{figure}

The orbit was converted to Galactic right-handed Cartesian XYZ
coordinates, with the convention that the Sun is centered at
$(X,Y,Z)=(-8,0,0)$ kpc and the direction of the X-axis points from the
Sun towards the Galactic center. Plots of the derived orbit projected
onto the Galactic $XY$ and $XZ$ planes are shown as solid (blue) lines
in Figure~\ref{fig:xy}. In these projections, it is clear that the
observed Virgo structure has just passed perigalacticon in its orbit
after swooping in toward the Galactic center from a distance of $>50$
kpc on a rather radial orbit. To give a sense of possible orbital
paths of debris due to both the uncertainties in our orbital
parameters and the dispersion of 3-D velocities in the VOD, we show 10
orbits generated by randomly picking from within $1\sigma$ Gaussian
distributions of the uncertainties on each of the orbital parameters
in Table~\ref{tab:stats}. Most of these ``perturbed'' orbits do not
depart drastically from the mean orbit.

These plots are similar in appearance to the orbits shown in
\citetalias{cgm+09a}. Our orbit, derived from 16 stars rather than a
single RR Lyrae, has similar eccentricity and apogalacticon as in the
previous work, confirming that we have measured kinematics of debris
from the same structure as the \citetalias{cgm+09a} study.  Note also
that on first glance our data support (within the uncertainties) an
association of NGC~2419 with the Virgo structure, as was suggested by
\citetalias{cgm+09a}. This and other plausible orbital associations
will be explored in the following section.

\subsection{Possible Orbital Associations}
\label{subsection:associations}

To search quantitatively for possible orbital associations of
satellites to our derived orbit, we compared the locations and
velocities of known GCs and dSphs with the path and velocity of the
mean orbit. This was done by defining a chi-squared value made up of
the sum of the residuals (scaled by the uncertainties) in position and
velocity between each point of the orbit and each Milky Way
satellite's known Galactic $XYZ$ position and $V_{\rm GSR}$
velocity. The uncertainties in spatial positions at each orbital
time-step were taken to be 30\%, and in $V_{\rm GSR}$ to be 30 km
s$^{-1}$. These large ranges should encompass the uncertainties in the
kinematics used to generate the mean orbit.\footnote{The assumption of a constant fractional uncertainty in the distance and fixed velocity uncertainty at all positions is chosen to simplify these calculations. In fact, for a given set of initial conditions and their associated uncertainties, the position and velocity uncertainties vary substantially at different points along the path of the orbit. Simulating this accurately would require a more sophisticated approach that is beyond the goals of this simple search for possible orbital associations.} The minimum $\chi^2$ is
located for each object, with the assumption being that this
represents the nearest point in phase space to the
satellite. Globular clusters within 15 kpc of the Galactic center were
excluded from the analysis because the Virgo substructure's orbit will
pass relatively near all of the globular clusters that are located
within this region. Without 3-D kinematics for all of these inner GCs,
it is difficult to assess their possible kinematical association with
the Virgo orbit.

The globular cluster NGC~2419, which was claimed by
\citetalias{cgm+09a} to be plausibly associated with the Virgo orbit,
is located $\sim30$ kpc from the apocenter of our orbit. Note,
however, that a slightly larger space velocity for Virgo debris at the
SA~103 position would send an orbit through almost exactly the
position of NGC~2419 (see, e.g., the maroon-colored ``perturbed'' orbit in
Figure~\ref{fig:xy}). The association of Virgo debris with this
globular cluster is thus tentatively confirmed by our data.

The lowest $\chi^2$ from our search (and, in fact, the only match with
$\chi^2 < 5.0$) was for ``SEGUE 1b'', which denotes the apparent
velocity substructure found by \citet{gws+09} and \citet{sgm+11} in
the field of view of the ultra-faint dSph SEGUE~1. SEGUE~1 itself is
spatially a good match to our Virgo substructure orbit, but the radial velocity
predicted by our orbit agrees much better with the debris at $\sim100$
km s$^{-1}$ higher velocities (and roughly the same distance,
according to \citealt{sgm+11}) than with SEGUE~1 itself.

In this section, we briefly discuss possible satellite associations
found by our fitting technique as well as other known stellar
overdensities; we will revisit each of these particular cases
one-by-one in the following section after we present the $N$-body
model based on our orbit.  The two main structures that were found to
be consistent with the integrated orbit were NGC~2419 and the velocity
substructure in the field of SEGUE~1 (which we denote ``SEGUE~1b'').
Another structure that may be related to the Virgo structure that is neither a GC
nor a dSph is the Pisces Overdensity.  This stellar overdensity is
apparently located near a turnaround point in its orbit, and possibly
contains distinct moving groups at positive and negative radial
velocity, which further suggests that it is debris near apocenter of
its orbit. An additional possibly-related structure is the debris
previously claimed
\citep{nyg+03} to be part of Sagittarius located at 90 kpc from the
Galactic center, in close proximity to NGC~2419. These structures are
briefly introduced here, and will be discussed in more detail in the
following section of this work.

\subsubsection{NGC 2419 / 90 kpc Sagittarius debris}

NGC 2419 is located at $(l,b) = (180\arcdeg,25\arcdeg)$ with
heliocentric and Galactocentric distances of 84 and 91 kpc
respectively. Its heliocentric radial velocity is $-20$ km s$^{-1}$,
corresponding to a Galactic standard of rest velocity of $-14$ km
s$^{-1}$ \citep{poa86}. \citet{nyg+03} find an excess of A-type stars
extending at least 10 degrees around NGC~2419 at about the same quoted
distance. The authors claim that since NGC~2419 is contained within
the debris, the tidal debris can be inferred to be at apogalacticon of
its orbit, consistent with their findings of being coincident with a
portion of the Sgr stream near its apogalacticon. However, this was
merely a tentative association of these stars with Sagittarius; our
current understanding of the Sgr tidal debris system (e.g.,
\citealt{lm10a}) suggests that there should not be debris at 90 kpc.

\subsubsection{SEGUE 1}

First discovered by \citet{bze+07} using SDSS imaging data, this dwarf
galaxy has subsequently been the subject of close examination because
of its peculiar properties, most notably its apparent high
mass-to-light ratio.  Located at (RA,Dec)$ = (152\arcdeg, 16\arcdeg)$
and $(l,b) = (220.5\arcdeg, 50.4\arcdeg)$ at a distance of $d_{\rm
hel} = 23 \pm 2$ kpc and $d_{\rm GC} = 28$ kpc, it is a good match to
our derived orbit of the Virgo substructure.  Belokurov et al. describe an apparent
tidal tail feature extending from the northeast to the southwest in
equatorial coordinates.  However, subsequent studies have not been
able to reproduce such results.  A study by \citet{gws+09} using
Keck/DEIMOS spectroscopy presented radial velocity and metallicity
data, and found a main grouping of stars at about $V_{\rm GSR} = 111 \pm
1.3$ km s$^{-1}$ with a smaller group of 4 stars at $\sim200$ km
s$^{-1}$.  They also obtain [Fe/H] $= -3.3 \pm 0.2$ for one of the
SEGUE 1 stars.  \citet{sgm+11} extended upon this survey using more
data and obtain similar numbers to those of Geha et al. They also
confirm the $\sim200$ km s$^{-1}$ feature from \citet{gws+09},
identifying 24 stars within the peak at $\langle V_{\rm GSR} \rangle =
203.8\pm1.7$ km s$^{-1}$, compared to the original 4 stars. These 24
stars represent a kinematically cold population, with $\sigma_V =
7.0\pm1.4$ km s$^{-1}$. \citet{sgm+11} suggest that the stars making
up this feature must be more metal-rich than those of SEGUE 1.

Our initial search for positional and kinematical associations to our
Virgo orbit found SEGUE 1 was a fairly good match. However, we noticed
that the $\sim200$ km s$^{-1}$ feature actually fits in better with
our findings at the given location. Thus the chi-square test was rerun
with an additional element, labeled ``SEGUE 1b'', to distinguish it
from the main SEGUE 1 population. The SEGUE 1b stars are an excellent
match to our orbit in all three spatial dimensions (within 1.3 kpc),
and offset from the best matching point by only $\sim10$ km s$^{-1}$
in radial velocity.

\subsubsection{The Pisces Overdensity}

The Pisces Overdensity was first found by \citet{sil+07} as an
overdensity (dubbed ``Structure~J'') of distant RR Lyrae stars in SDSS
Stripe 82. It was subsequently examined by \citet{web+09} and shown to
stretch across an area from $63\arcdeg < l < 93\arcdeg$ and
$-60\arcdeg <b< -46\arcdeg$ at a Galactocentric distance of $79.4 \pm
14.1$ kpc and a metallicity of [Fe/H]$ = -1.48 \pm 0.28$.
\citet{kgs+09} measured the radial velocities of RR Lyraes within SDSS
Stripe 82 and found two distinct groups of stars, one centered at a
heliocentric radial velocity of $-75$ km s$^{-1}$ and the other around
$-198 < V_r < -155$ km s$^{-1}$.  They find similar distances and
metallicities as Watkins et al.  Later, \citet{sig+10} expanded the
selection of stars of Kollmeier et al. and confirmed two distinct
moving groups of Galactocentric velocity of $-52 \pm 11$ km s$^{-1}$
and $50 \pm 3 $ km s$^{-1}$.  This finding of two separate velocity
groups in opposite directions suggests that this overdensity is at a
turning point in its orbit.  This too agrees nicely with the orbit
that we have obtained in this area. \citet{sjm+10} find a structure
among 2MASS-selected M giants in the vicinity of the quoted location
of the Pisces Overdensity, which is labeled ``A16'' in their paper.
They find that the peak density lies just outside the range of SDSS
Stripe 82. They hypothesize that the Pisces structure is an extended
remnant of a satellite that was on a highly eccentric orbit and got
tidally disrupted.  However, in the same paper, they were not able to
detect VSS members, calling into question such an association with
VSS.

\section{Discussion: the Progenitor of the Virgo Stellar Substructure}

\subsection{N-body Simulation}

With the orbital parameters of Virgo debris well-constrained, we turn
to the question of what type of progenitor could produce a structure
with properties like those of the Virgo substructure. With the extant
spatial and kinematical data, it is unclear whether all of the
numerous detections of substructure in Virgo are related. However, we
will show that a simple model based on our precise kinematics explains nearly all of the observed properties.

To understand the orbital dynamics of the Virgo substructure
progenitor more clearly, semi-analytical $N$-body simulations were
generated using our derived orbit. The kinematical parameters
determined for Virgo debris in SA~103 were used to generate orbital
models using the same gravitational potential as the test particle
orbit. The simulation templates used were 10,000 particle Plummer
spheres placed on our derived orbit for the Virgo structure.  The
satellites were evolved in the Galactic potential for 3-4 Gyr using
the gyrfalcON tool of the NEMO Stellar Dynamics Toolbox
(\citealt{d02}), and their ending time tailored to place the core of
the progenitor near the current VOD. This strategy of placing the
progenitor (or its remains) at the position of the VOD is motivated by
our derived orbit. Pile-ups of tidal debris are typically located at
the apo- or peri-center of a satellite's orbit; however, we showed in
the previous section that our measured kinematics place Virgo
substructure stars at an orbital phase just past pericenter. Thus, in
order to have a cloud-like pile-up of stars at the VOD position (i.e.,
at an orbital phase that is {\it not} at peri- or apo-center), the
stellar overdensity is likely the remains of the progenitor itself.

\begin{figure}[!Ht]
\epsscale{0.7}
\plotone{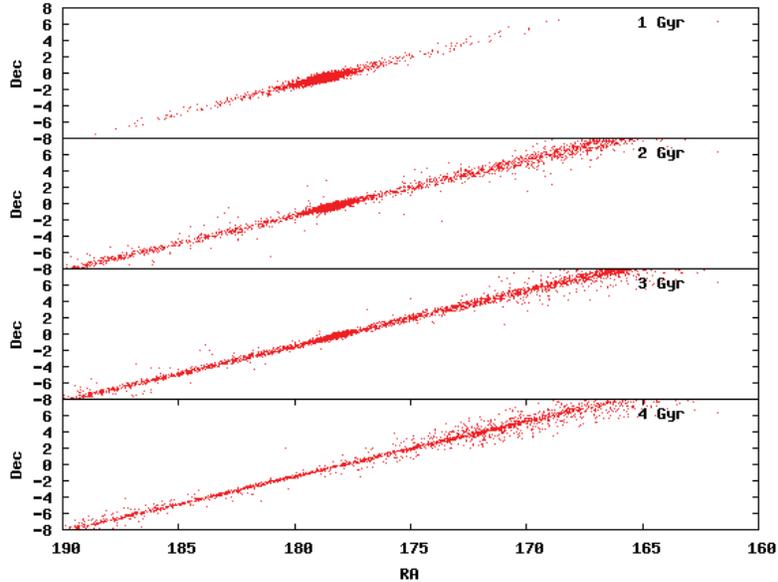}
\caption{Shown here is an $N$-body simulation of a globular cluster sized object evolved along the derived test particle orbit, in 4 different time slices, spaced 1~Gyr apart. The cluster is modeled as a Plummer sphere with mass of
$M=1\times10^5 M_{\Sun}$ and Plummer scale radius of 50 pc, represented by
10,000 red particles of equal mass. The object is seen to rapidly
stretch out into a thin filament, falling apart between 3-4 Gyr, or
after 3 perigalactic passages.}
\label{fig:glob}
\end{figure}

Because the progenitor of the Virgo structure is unknown (if it even
still exists), we must explore the behavior of disrupting satellites
of different masses. Initially, we placed a globular-cluster-sized
object with $M=1\times10^5 M_{\Sun}$ and Plummer radius $r_{\rm pl} =
50$ pc on our derived orbit.  On the eccentric orbit derived for Virgo
stars, the globular cluster is rapidly shredded upon its pericentric
passage. The $N$-body results of this globular cluster progenitor are
seen in Figure~\ref{fig:glob}, which focuses on the region of sky
around SA~103. The satellite is already extended into a narrow tidal
stream after its first perigalactic passage (in the upper panel), and
is completely disrupted by around 3 to 4 Gyr, corresponding to the
third perigalactic passage.  This creates a narrow, well defined
stream in the sky, contrary to observations of a large cloud-like
feature seen in Virgo. We note also that remnants of disrupted
globular clusters are typically rather kinematically cold; the absence
of a narrow velocity peak among the Virgo substructure data thus also
argues against a globular cluster progenitor. Therefore, a globular
cluster-like object is unlikely to be the progenitor of the
substructure(s) in Virgo.

\begin{figure}[!Ht]
\epsscale{0.7}
\plotone{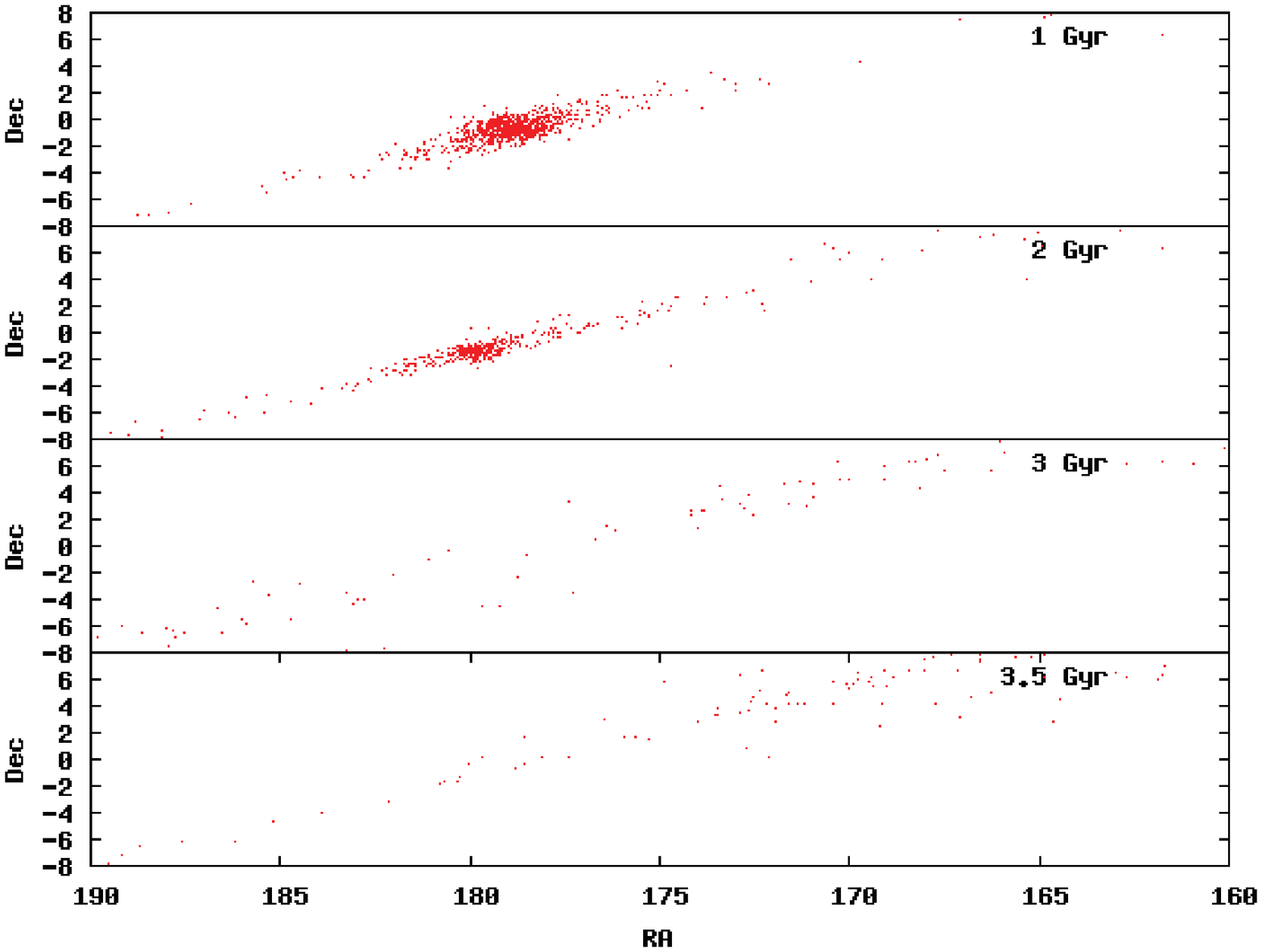}
\caption{Results of an $N$-body simulation of an Orphan-like progenitor
(a dwarf-galaxy-sized object) evolved along our derived Virgo
structure orbit. The panels show snapshots at the position of SA~103
in four time slices spaced $\sim$1~Gyr apart. This object was
configured as a Plummer sphere with mass of $M=3\times10^6 M_{\Sun}$
and a scale radius of 200 pc. The satellite dissolves quickly (in less
than 3~Gyr), with the remnant showing a narrow stream-like structure
at early times and virtually no spatial concentration at later
stages.}
\label{fig:orphan}
\end{figure}

Having ruled out a globular cluster as the Virgo substructure progenitor, we then
tried a larger, more massive dwarf galaxy-like
satellite. Specifically, we first simulated a dSph of mass similar to
that derived by \citet{nwy+10} for the Orphan stream progenitor
($M=3\times10^6 M_{\Sun}$), configured in a Plummer sphere of radius
$r_{\rm pl} = 0.2$ kpc. This low-mass dwarf galaxy was run through an
$N$-body simulation with the same MW parameters as the globular
cluster model, on our derived Virgo substructure orbit. As with the
globular cluster, the Orphan-like progenitor disrupts quickly on this
destructive orbit. As can be seen in Figure~\ref{fig:orphan}, this
comparatively kinematically ``warmer'' object dissolves much more
rapidly than the globular cluster. The remnant becomes more ``puffy''
than the disrupted globular cluster, but still remains embedded in a
narrow stream. This more closely resembles what is seen in Virgo, and
suggests that an even larger, more massive dSph may produce a
large-area, cloud-like overdensity when disrupted on this orbit.

\begin{figure}[!Ht]
\epsscale{1.0}
\plotone{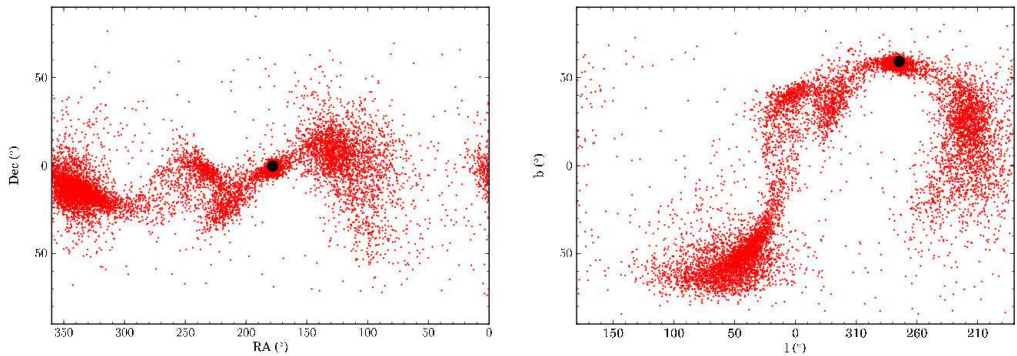}
\caption{The results of an $N$-body simulation of a Sagittarius-sized ($M=1\times10^9 M_{\Sun}, r_{\rm pl}=0.9$ kpc) Plummer sphere on our derived Virgo substructure orbit. The left panel depicts the debris in
equatorial coordinates and the right panel shows the simulation
results in Galactic coordinates.  The black point represents the
present location of the observed Virgo debris in SA~103. This $N$-body
simulation was initially placed 3 Gyr in the past and integrated
forwards for 2.95 Gyr. Three overdensities are immediately apparent:
The ones at $l=60\arcdeg$ and $l=210\arcdeg$ are due to piling up of
debris at apogalacticon. The third overdensity located at
$l=270\arcdeg$ is progenitor related; in other words, the location of
this cloud of debris changes depending on the duration of the
simulation. We tailored the run-time of this particular simulation to
place the progenitor (or its remains) at this position.}
\label{fig:sgrlb}
\end{figure}

\begin{figure}[!Ht]
\plotone{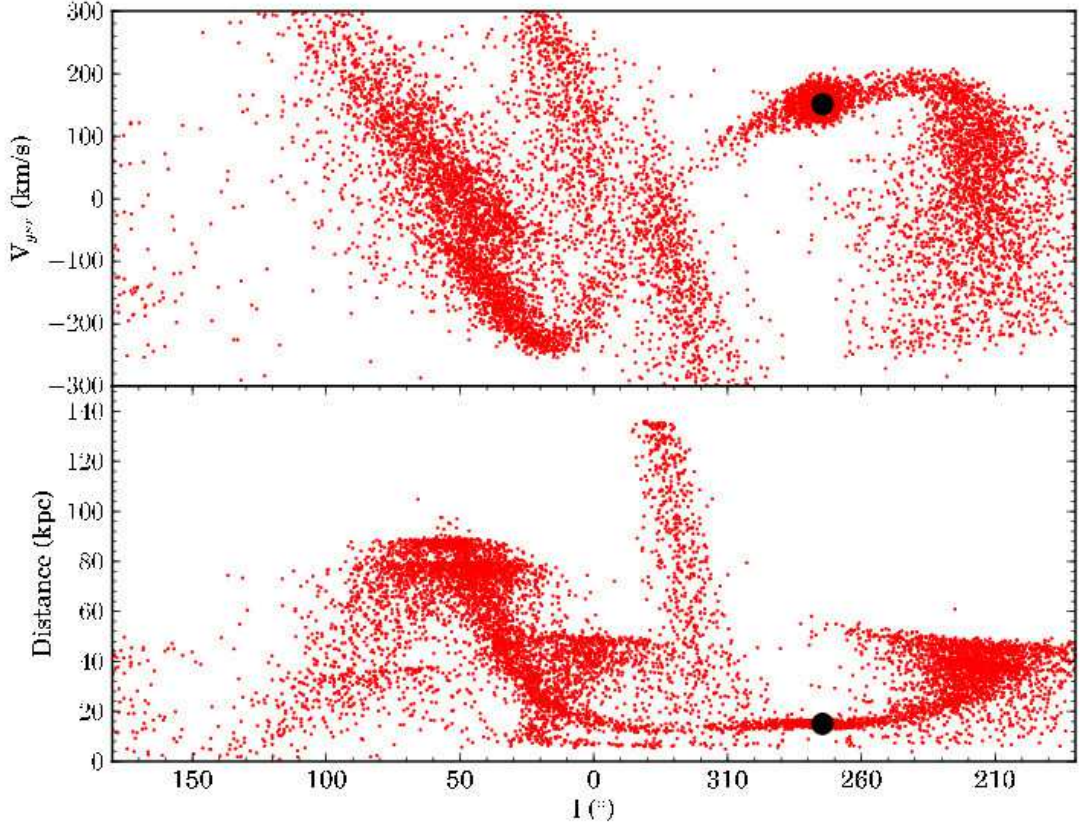}
\caption{$N$-body results from the simulation with a Sagittarius-like progenitor, as in Figure~\ref{fig:sgrlb}. The upper panel shows $V_{\rm GSR}$ vs Galactic longitude while the lower
panel shows heliocentric distance vs. Galactic longitude. The model runtime was tailored to place the progenitor (and/or its remains) at the position of SA~103 (the large black dot).}
\label{fig:sgrvd}
\end{figure}

\begin{figure}[!Ht]
\includegraphics[width=3.5in,trim=0.2in 0.2in 0.2in 0.2in,clip]{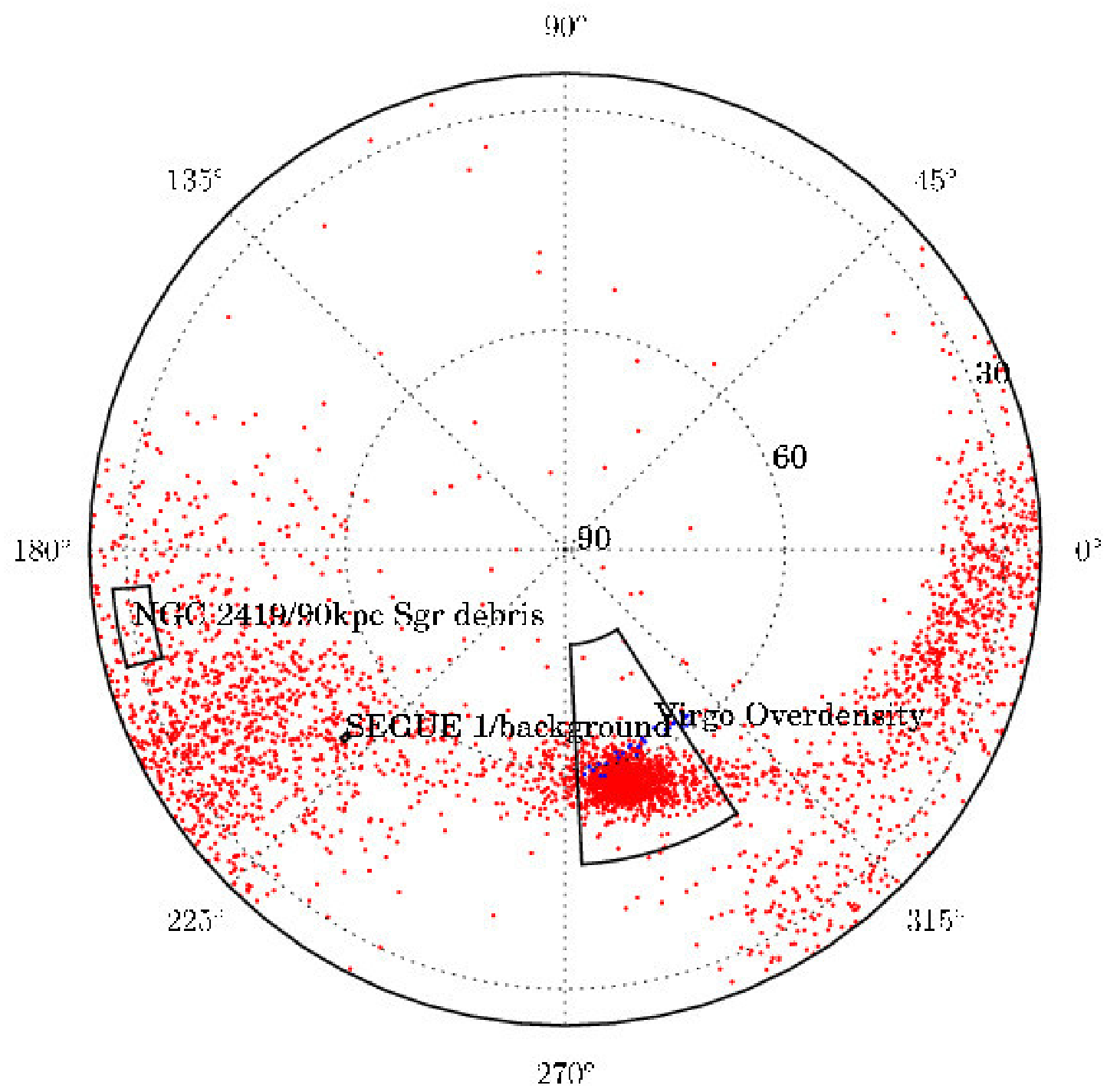}
\includegraphics[width=2.8in]{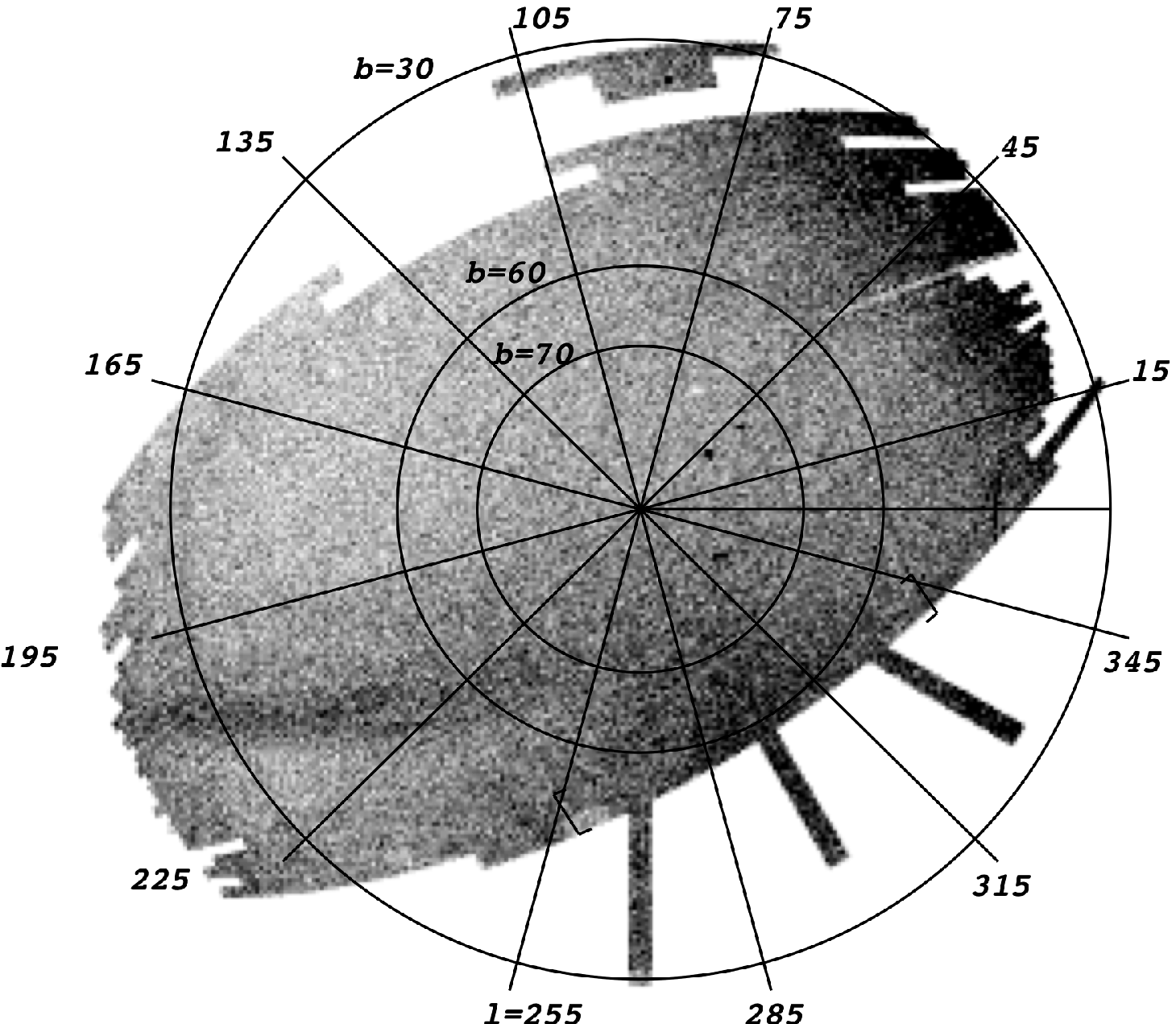}
\caption{Left panel: the Sagittarius-sized $N$-body simulation in
  polar Galactic coordinates, centered at the North Galactic Cap. Here
  an overdense region around $(l,b)=(285\arcdeg,55\arcdeg)$ is clearly
  visible. Approximate boundaries of the possible associations are
  also drawn in the plot.  Right panel: Density map of F-turnoff stars
  in the Northern Galactic Cap from Newberg et al. 2007. These stars
  were obtained from SDSS DR5 based on the criteria $0.2<(g-r)_0<0.3$,
  $(u-g)_0 > 0.4$, and $20.0<g_0<21.0$. Darker areas contain more F
  turnoff stars. Comparing this figure to the left panel, one can see
  an apparent agreement between the observed location of the Virgo
  debris and that predicted by the N-body model. The Virgo overdensity
  is visible at the lower right edge of the data in the right panel.
}
\label{fig:sgrlbpolar}
\end{figure}

\begin{figure}[!Ht]
\epsscale{0.6}
\plotone{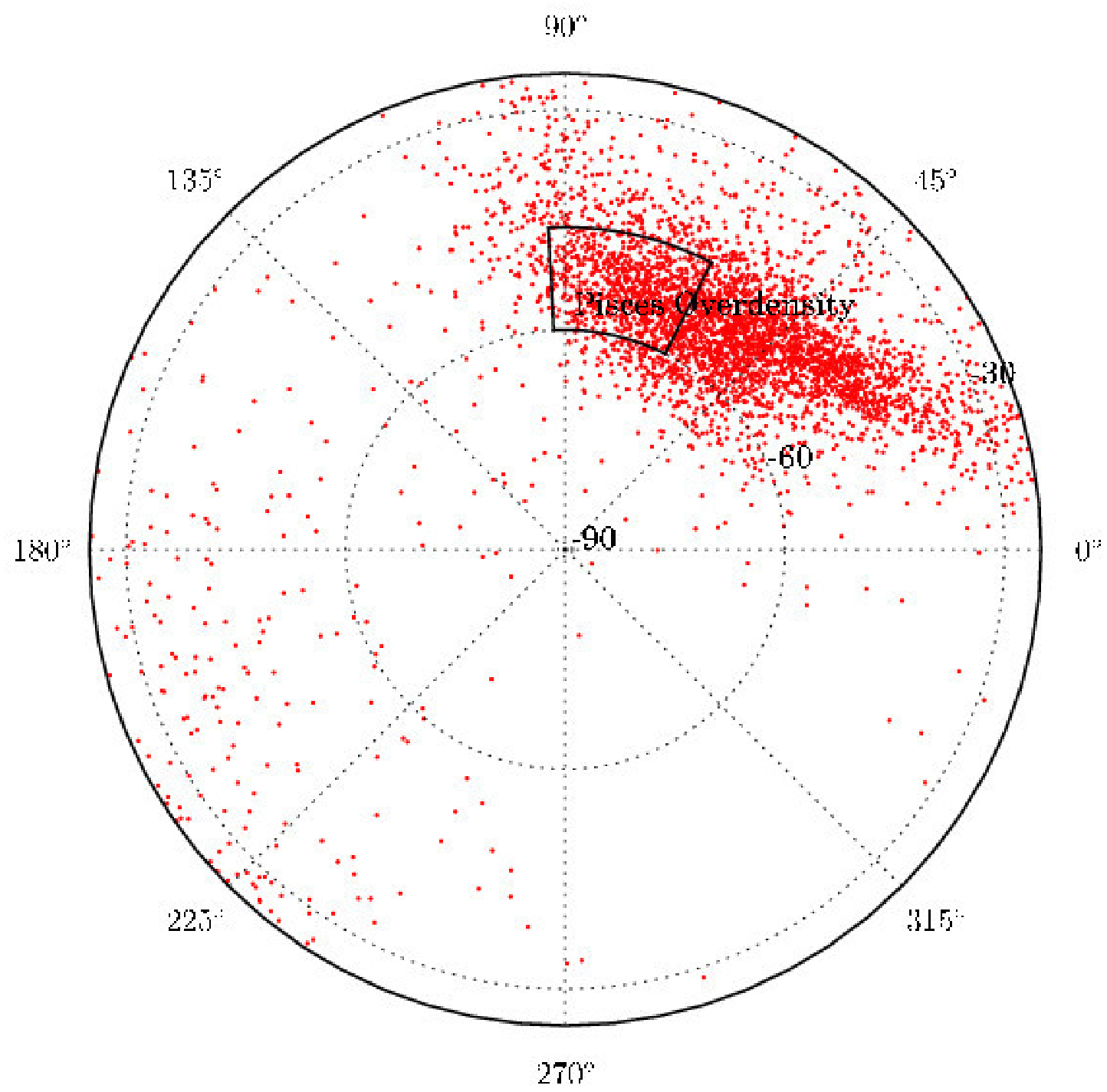}
\caption{$N$-body model debris in the southern Galactic hemisphere. An
overdensity of model debris is seen that is nearly coincident with the Pisces Overdensity between $60 \lesssim l \lesssim 90\arcdeg$. These stars, located at distances of $\sim80-90$ kpc, correspond to the apogalacticon of our derived orbit. }
\label{fig:sgrlbpolarsouth}
\end{figure}

Finally, we simulated a 10,000 particle, Sagittarius-sized template
dwarf galaxy (a Plummer sphere with mass of $M \sim 10^9~M_{\Sun}$ and
scale radius $r_{\rm pl} = 0.9$ kpc) on the Virgo test-particle orbit.
The results of this gravitational $N$-body simulation are displayed in
Figures~\ref{fig:sgrlb} (spatial distribution), \ref{fig:sgrvd}
($V_{\rm GSR}$ and distance vs. longitude), and \ref{fig:sgrlbpolar}
(polar projection of the spatial positions). The disruption of a
Sgr-sized object on this orbit produces several structures that span
large (hundreds of square degree) areas of the sky
(Figure~\ref{fig:sgrlb}). The large, diffuse overdensities centered at
$(l, b) \sim (60\arcdeg, -50\arcdeg) [(\alpha, \delta) \sim
(330\arcdeg, -5\arcdeg)$] and $(l, b) \sim (210\arcdeg, 40\arcdeg)
[(\alpha, \delta) \sim (110\arcdeg, 20\arcdeg)$] are the results of
tidal debris piling up at apogalacticon. These distant debris (at
$R_{\rm GC}\sim80$ and $\sim50$ kpc, respectively; see
Figure~\ref{fig:sgrvd}) piling up at the turning points of the
disrupting dSph's orbit are expected for an object on such a radial,
destructive orbit (see, e.g., the ``umbrella-like'' structures in
\citealt{jbs+08}). One may expect a dSph seen shortly after a
perigalactic passage to have a distorted morphology with well-defined
tidal tails emanating from it. However, the $N$-body model guided by
our precise kinematics of Virgo stars shows that the large tidal shock
imparted to a Sgr-like progenitor upon its passage $<10$ kpc from the
Galactic center causes the dSph to inflate to an apparent size (at $d
= 15$ kpc) of hundreds of square degrees (note that we are not
concerning ourselves about whether the particles are bound to the
progenitor at this stage; most of the stars must be unbound at this
point in the object's evolution). This is seen more clearly in the
left panel of Figure~\ref{fig:sgrlbpolar}, which shows a polar plot of
$N$-body model debris in Galactic coordinates, centered on the north
Galactic pole. Given for comparison in the right panel of
Figure~\ref{fig:sgrlbpolar} is a reproduction of Figure 5 from
\citet{nyc+07}, in which the Virgo substructure is clearly seen as an
overdensity of photometrically-selected F-turnoff stars from SDSS. The
progenitor (or its remains) in our model spans roughly the same area
as the overdensity seen by Newberg et al., centered on the same
position, including possibly an elongated shape along Galactic
latitude (i.e., from left to right in the figure).

This structure, seen at roughly the position of SA 103 (denoted by the
large black dot in Figure~\ref{fig:sgrlb} at $(l, b) \sim (275\arcdeg,
59\arcdeg)$, or $(\alpha, \delta) \sim (179\arcdeg, -1\arcdeg)$) and
spanning $\sim500$ square degrees, contains $\sim16\%$ of the original
10,000 particles making up the progenitor. Since the progenitor had a
mass of $10^9 M_{\Sun}$, this corresponds to a mass of
$\sim1.6\times10^8 M_{\Sun}$ in the area surrounding the remains of
the progenitor. As an exercise, we estimate the luminosity
corresponding to this mass by assuming a mass-to-light ratio typical
of Milky Way dSphs,\footnote{In our model, the object is probably no
longer a bound system, making the assumptions of these calculations
dubious. We present these results simply as an attempt to compare
$N$-body particles to the observed stellar overdensity.}
($\frac{M}{L})_{\Sun} = 100$ (see, e.g., \citealt{wmb+10},
\citealt{wmo+09}, \citealt{sg07}); this yields a $V$-band absolute
magnitude of $M_{\rm V} = -10.7$ (or $M_{\rm V} = -13.2, -8.2$,
respectively, for $\frac{M}{L}_{\Sun} = 10, 1000$). This compares
favorably to the estimate of the total integrated light of the VOD by
\citet{pdk+09} in a 760 deg$^2$ area of $M_{\rm V} = -11.9$ (or
$M_{\rm V} = -10.1$ using the more nearby
\citealt{jib+08} distances). \citet{jib+08} found a total $r-$band
luminosity of $0.1\times10^6 L_{\Sun}$, or $M_r = -7.8$, for the
overdensity assuming a distance of 10 kpc and 1000 deg$^2$ area. Thus,
the remains of the Virgo progenitor as determined by our derived orbit
are consistent with the observed stellar overdensity that covers
hundreds of square degrees.

Another mystery surrounding the numerous overdensities detected in
Virgo is the disagreement among radial velocities between studies of
debris at different positions. As we showed in Section
\ref{section:candidates}, even in SA~103, which spans only
$40\arcmin\times40\arcmin$, there is a significant excess of
high-$V_{\rm GSR}$ stars that forms a broad peak (see
Figure~\ref{fig:SA103histvgsr}). This is more obvious among SDSS RVs
shown in Figure~\ref{fig:DR8histvgsr} -- a broad excess relative to
the Besan\c{c}on model is apparent at $100 \lesssim V_{\rm GSR}
\lesssim 250$ km s$^{-1}$. This is much too broad to be the signature
of a kinematically-cold tidal stream. If the VOD were due to the piling
up of tidal debris at the apocenter of a satellite orbit, then a large
velocity dispersion could be expected; however, the mean velocity at
such a turnaround point must be $V_{\rm GSR} \sim 0$ km s$^{-1}$ rather
than $\sim150$ km s$^{-1}$ as observed. Furthermore, we have shown
that stars in the Virgo substructure are not at apogalacticon, having rather just
passed their {\it peri-}galacticon. Examination of the upper panel of
Figure~\ref{fig:sgrvd}, however, shows that our $N$-body model offers
an explanation for the large spread in velocity of Virgo debris. The
progenitor, having just experienced a close encounter with the
Galactic center, has been kinematically heated so that its velocity
dispersion has increased to $\sim20$ km s$^{-1}$ or more and its RVs span
$\sim100$ km s$^{-1}$. Note that the model seen in Figure~\ref{fig:sgrvd} does not predict debris with $V_{\rm GSR} > 200$ km s$^{-1}$ near the progenitor, as is seen in both the SA~103 data (Figure~\ref{fig:SA103histvgsr}) and SDSS (Figure~\ref{fig:DR8histvgsr}) as a long tail to high velocities. It is possible that additional substructure is present in this region of the sky. Alternatively, this long tail may be a feature of the disrupting Virgo progenitor  that our simple model has not addressed -- further modeling exploring the progenitor's mass and mass distribution may be able to reproduce this feature.

To summarize, our $N$-body model of a Sagittarius-sized dwarf galaxy
with mean kinematics of the 16 stars we determined to be Virgo
substructure members qualitatively reproduces many of the observed properties of the
stellar excess in Virgo. We have shown that a large, cloud-like stellar structure like that in Virgo is likely to have originated in the tidal disruption of a massive dwarf galaxy. No dwarf galaxies are found along the integrated path of the orbit we have derived, leading us to conclude that the Virgo substructure is itself the remains of the progenitor. The slight mismatch in the position of the disrupted progenitor in our model and that of the observed overdensity in Virgo is a consequence of the model runtime and our lack of knowledge of the central position of the stellar overdensity, which may lie outside the observed SDSS footprint. Details such as the size of the overdensity and the direction of its elongation may also be slightly discrepant between the model and observations, but are in rough qualitative agreement. Many of these differences are likely due to the fact that the initial conditions used to generate the model -- namely, the kinematics of debris in SA~103 -- may not actually correspond to the center of the progenitor. A more comprehensive exploration of parameter space (including, but not limited to, the mass of the progenitor, its position, and its initial density configuration) could likely resolve some of these issues, but is beyond the scope of this work. It is clear, however, that the morphology, spatial extent,
luminosity, and large velocity dispersion of the observations are all qualitatively consistent with our simple 
model. We thus suggest that a massive dwarf galaxy was the progenitor
of the substructure in Virgo.

\subsection{Other Possible Progenitors or Associated Overdensities}

In Section \ref{subsection:orbit} we discussed a search for Milky Way
satellites and stellar overdensities that may be associated with the
Virgo substructure. While initially the search was conducted in hopes
of identifying the progenitor of the VOD, our modeling results suggest
that the VOD itself is likely the remains of the progenitor. However,
the infalling dwarf spheroidal that produced the VOD may have had
satellites of its own, similar to the globular clusters associated
with the Sgr and Fornax dSphs. If the progenitor had globular clusters
orbiting in its potential, these should be found along the orbit of
the progenitor. Also, since the VOD is near the pericenter of the
progenitor's rather destructive orbit, there should be additional
tidal debris that has collected at the apogalactica of the orbit, as
seen in Figures~\ref{fig:sgrlb} and \ref{fig:sgrvd}. In this
subsection, we revisit the satellites and stellar substructures we
identified in Section \ref{subsection:associations}, and discuss them
in the context of the $N$-body model results.

\subsubsection{NGC 2419 and Nearby 90-kpc Stellar Overdensity}

The distant Milky Way globular cluster NGC 2419 was suggested by CD09
to be plausibly associated with their orbit of the VSS based on a
single RR Lyrae star. The orbit passes near NGC 2419 at the next
apocenter ahead of the Virgo substructure along its orbit. The cluster
has a radial velocity of $\langle V_{\rm GSR} \rangle = -14$ km
s$^{-1}$, which, because it is close to zero, is consistent with
NGC~2419 being at the turnaround point of its orbit. We find that our
orbit does not reach the $\sim80$ kpc distance of NGC~2419, instead
turning around at $\sim50$ kpc at its nearest apocenter. However,
within the uncertainties on the orbital parameters (as illustrated by
the different orbits in Figure~\ref{fig:xy} with kinematics selected
from within the $1-\sigma$ uncertainties), the orbit could plausibly
be associated with a satellite or debris at such distances.  NGC~2419
is an unusually large, luminous globular cluster, which has led to
speculation that it is actually the stripped core of a former dwarf
spheroidal galaxy (e.g., \citealt{vm04,mv05,cks+10,chk11}). One might
thus suggest NGC~2419 to be the progenitor of the Virgo
substructure. We believe this is unlikely, because NGC~2419 has a
metallicity of [Fe/H] = -2.1 \citep{skk88}, making it more metal-poor
than the debris we have studied in SA~103, and even more deficient
than the RR Lyrae studies have found for VOD/VSS stars. Because
metallicity gradients in dwarf galaxies lead to more metal-rich
populations being concentrated at their cores, tidal streams from
disrupted dSphs should have mean metallicities that decrease with
distance along the stream from the progenitor
(e.g. \citealt{cmc+07}). It is hard to imagine how, if NGC 2419 is the
progenitor, it could have a mean metallicity lower than that of all
known debris. It is also possible that the cluster fell in with the
(dwarf-galaxy) progenitor. However, we note that \citet{v07a} pointed
out that the globular clusters associated with dwarf galaxies are
faint relative to the general globular cluster population; this would
appear to rule out the extremely luminous NGC~2419 as a
globular-cluster companion of an infalling dwarf galaxy.

Furthermore, \citet{nyg+03} identified an overdensity of A-type
(primarily BHB) stars at a mean distance of 83 kpc spanning over 200
deg$^2$ in the region surrounding NGC~2419. Very few BHB stars should
be present that far out in the halo, and it was originally suggested
by Newberg et al. that these stars are Sagittarius tidal debris. In
our $N$-body model of the Virgo progenitor, the position of the
\citet{nyg+03} stars on the sky is consistent with the collection of
Virgo-structure stars at the apocenter of its orbit. We note that the
$N$-body model produces debris about 40\% closer than either NGC~2419
or the Newberg et al. stars, but this is not troubling. The orbit
traverses distances consistent with both NGC~2419 and the A-star
overdensity, and we have seen that $N$-body models using slightly
different kinematics (from within our error distribution) would
produce model debris at those distances as well. We thus suggest that
NGC~2419 and the ``cloud'' of A stars nearby it may be associated with
the Virgo progenitor.

\subsubsection{Substructure in SEGUE 1 Field of View}

Another of the close matches to our orbit is the dwarf galaxy SEGUE 1,
which lies directly along the path of the Virgo structure
orbit. However, we are not suggesting here that the extremely
metal-poor, dark matter-dominated SEGUE 1 dSph is associated with the
Virgo structure.  Instead, it is the ``extra'' radial velocity peak
identified by
\citet[with 4 stars]{gws+09}, later seen more strongly by \citet[8
stars]{nwg+10}, and finally detected with 24 stars by \citet{sgm+11}
that we believe to be tidal debris from the Virgo system. Because the
spectroscopic targets were selected by Simon et al. to be consistent
with the SEGUE 1 distance ($23\pm2$ kpc), the stars in the additional
RV peak are presumably at roughly this same distance (\citealt{sgm+11}
quote a distance of $\approx22$ kpc). However, their mean velocity is
at $V_{\rm GSR}=204$~km~s$^{-1}$, offset by $\sim100$~km~s$^{-1}$ from
the SEGUE 1 velocity. Our $N$-body model predicts Virgo debris should
be present at this position ($l,b = 220, 50\arcdeg$) at a distance of
$\sim25$ kpc (see Figures~\ref{fig:sgrlb} and \ref{fig:sgrvd}).
Furthermore, the upper panel of Figure~\ref{fig:sgrvd} shows that the
debris in our model is at $150 < V_{\rm GSR} < 200$ km s$^{-1}$, or
almost exactly where the 24 ``excess'' stars in the \citet{sgm+11}
study appear. We therefore conclude that the debris in the SEGUE 1
field of view is plausibly associated with the Virgo stream system. We
note also that the mean metallicity is suggested by \citet{sgm+11} to
be [Fe/H] = -1.3 for these 24 stars in the ``300 km s$^{-1}$ stream''.
This is consistent with the metallicity we have measured for debris in
SA 103. Lending further credence to our claim of an association with
the stream from the Virgo progenitor is the fact that both
\citet{sgm+11} and Niederste-Ostholt et al. (2009) see an east-west
extension of the 300 km s$^{-1}$ stream stars. In fact, Figure 8b of
\citet{sgm+11} shows the stars in this velocity feature to possibly be
oriented along a southeast-to-northwest direction, which is exactly
the orientation of the debris from our model of the Virgo progenitor.
Based on all these facts taken together, we believe that the 300 km
s$^{-1}$ stream in the SEGUE 1 field of view is debris associated with
the Virgo substructure.

\subsubsection{Pisces Overdensity}

Finally, we note that the ``Pisces Overdensity'', a stellar excess
spanning a large area on the sky at a Galactocentric distance of
$79\pm14$ kpc \citep{sil+07,web+09}, is located at the position and
distance of the most recent apocenter of our Virgo orbit. This
overdensity was first found among RR Lyrae in SDSS Stripe 82 data at
$(\alpha,\delta) \sim (355\arcdeg,0\arcdeg)$, or $(l,b) \sim
(88\arcdeg,-58\arcdeg)$, and subsequently seen to stretch to $(l,b)
\sim (105\arcdeg,-53\arcdeg)$ in 2MASS-selected M-giant stars
\citep{sjm+10}. Examination of Figures~\ref{fig:sgrlb} and \ref{fig:sgrvd}
shows a pile-up of model debris at the position and near the distance roughly
corresponding to the Pisces Overdensity. This is also illustrated in
the polar view of model debris in the southern Galactic cap seen in
Figure~\ref{fig:sgrlbpolarsouth}, which overplots the approximate
location of the Pisces Overdensity atop our model. Because this
substructure has been detected in metal-poor RR Lyrae
\citep{sil+07,web+09} as well as relatively metal-rich 2MASS-selected
M-giant stars \citep{sjm+10}, it has already been suggested to be
related to a dwarf-galaxy disruption event. It is therefore remarkable
that our $N$-body model for a disrupting dSph on the Virgo orbit
predicts just such a ``cloud'' of tidal debris as has been
observed. The radial velocity one would expect at the turnaround point
of the orbit is $V_{\rm GSR}=0$ km s$^{-1}$, with some large spread
about zero velocity due to the viewing angle roughly along the path of
both the outward-streaming and inward-bound debris (see
Figure~\ref{fig:sgrvd} for an illustration of the large range of
expected $V_{\rm GSR}$ in a $\sim20\arcdeg$ range of Galactic
longitude). Spectroscopy of RR Lyrae in the Pisces structure by
\citet{kgs+09} was combined with additional spectra by
\citet{svd+10} to arrive at a sample of 12 RR Lyrae in the
overdensity. These 12 stars separate into two distinct peaks at
$\langle V_{\rm GSR} \rangle =50\pm3$ km s$^{-1}$ with $\sigma=10\pm3$
km s$^{-1}$, and $\langle V_{\rm GSR} \rangle =-52\pm11$ km s$^{-1}$,
$\sigma=23\pm9$ km s$^{-1}$. \citet{svd+10} considered the possibility
that the two groups are drawn from the same Gaussian distribution, and
concluded that the hypothesis of normality can be rejected at the
$\sim95\%$ level. However, from the small number of stars, it is
unclear whether the two RV peaks correspond to approaching and
receding portions of the same tidal stream, or separate
substructures. However, it is reasonable to conclude that either (or
perhaps both) of these velocities are consistent with debris from the
Virgo progenitor as modeled by us. We note, though, that
\citet{svd+10} find that although the two velocity groups are centered
at roughly the same positions, one of the stars in the
negative-velocity group extends to lower right ascension. Similarly,
in Figures 12 and 14 of \citet{web+09} the Pisces Overdensity appears
to be closer at lower RA (and lower Galactic longitude) than at its
peak. Both of these findings are consistent with our tidal stream
model, wherein the debris on the lower-longitude side of the
apoGalacticon is on the approaching portion of the stream. We conclude
that it is likely that the Pisces Overdensity has its origin in the
disruption of the Virgo progenitor. However, further observations and
detailed modeling of both the Virgo and Pisces structures are needed to
confirm this.

\section{Conclusion}

We have presented 3-D kinematics of candidate Virgo substructure
members in SA~103, a $40\arcmin \times 40\arcmin$ field located at
($\alpha, \delta)_{\rm 2000} = (178.75\arcdeg, -0.59\arcdeg)$, or $(l,
b) = (274.6\arcdeg, 59.1\arcdeg)$. This work builds upon that of
\citetalias{cgm+09a}, in which an orbit was presented for a single VSS
RR~Lyrae candidate from the SA~103 data set. We have obtained
follow-up spectra of a total of 215 stars from the Kapteyn's Selected
Areas proper motion survey outlined in \citet{cmg+06}. From the
derived radial velocities, we isolate an excess of relatively
high-velocity stars that are unexpected from smooth Milky Way
populations. These stars, at $100
\lesssim V_{\rm GSR} \lesssim 350$ km s$^{-1}$, do not make up a
kinematically cold ``peak'' in the RVs, but rather span nearly the
entire high-velocity range. However, we show that the proper motions
of most of the stars from these high velocities are consistent with
their being kinematically associated. Furthermore, these high-velocity
stars inhabit regions of the color-magnitude diagram that place them
all at nearly the same distance, $\sim14\pm3$ kpc, from the Sun.

We identify a total of 16 candidate Virgo substructure members, from
which we derive a weighted mean radial velocity $V_{\rm GSR} =
152.6\pm22.0$ km s$^{-1}$ and proper motion $(\mu_\alpha \cos{\delta},
\mu_\delta) = (-5.24, -0.91)\pm(0.43, 0.46)$ mas yr$^{-1}$. The mean metallicity of the substructure is estimated to be [Fe/H] $\sim$ -1.1. From the
mean kinematics and the derived distance of $14\pm3$ kpc, we generate
an orbit in a model Galactic potential. The Virgo substructure stars
are on an eccentric orbit ($e = 0.81^{+0.10}_{-0.27}$) with peri- and
apo-centers of $R_{\rm p} = 5.6^{+0.9}_{-0.8}$~kpc and $R_{\rm a} =
52^{+87}_{-24}$~kpc, respectively.

We searched for satellites or known overdensities that can plausibly
be associated with the Virgo progenitor. Based on the positions and
radial velocities of the derived orbit, we find that known tidal
debris in the field of view of the ultra-faint dwarf SEGUE~I match our
predicted orbit in position and velocity. While it is intriguing to
find such a close match to the predicted path and kinematics of the
Virgo progenitor orbit, one must explain the fact that the 24 stars
making up the ``300~km~s$^{-1}$ debris'' near SEGUE~I were found by
\citet{sgm+11} to have a velocity dispersion of only
7.0~km~s$^{-1}$. It is possible that these stars near SEGUE~I, as well
as many of the apparently cold stellar substructures in Virgo, are
kinematically cold sub-clumps within the Virgo debris. Another
plausible match to our derived orbit is the globular cluster
NGC~2419. This distant cluster was suggested by
\citetalias{cgm+09a} to be associated with the Virgo substructure,
possibly as its progenitor. It is difficult to explain how the
metal-poor ([Fe/H] = -2.1) NGC~2419 could have been the origin of the
more metal-rich Virgo debris; rather, it is more likely that NGC~2419
fell into the Milky Way with the Virgo progenitor. We also note
that there is a known overdensity of A-stars found by \citet{nyg+03}
to span $\sim200$ deg$^2$ around NGC~2419 at a mean distance of 83
kpc. This cloud of stars may be associated with NGC~2419, the Virgo
progenitor, or both.

Through the use of $N$-body models of satellites on this rather
destructive orbit, we show that the large spatial extent of the Virgo
substructure does not arise if the progenitor of the stellar
overdensity was a globular cluster or low-mass dwarf galaxy. Rather,
the progenitor of the Virgo structure must have been a massive,
Sagittarius-like dwarf spheroidal galaxy of mass $\sim 10^9
M_{\sun}$. The $N$-body model of a Sgr-mass dSph on our derived orbit
qualitatively reproduces the observed spatial extent, velocity spread,
and estimated stellar mass of the Virgo overdensity. We thus conclude
that the entire cloud-like Virgo substructure is likely the tidal
debris remnant from a recently-disrupted dwarf galaxy. Our model also
produces a ``cloud'' of debris near the position and distance of
NGC~2419 (and the surrounding A-star overdensity), corresponding to
the apocenter of the orbit. Another diffuse overdensity is predicted
by the model near the location of the Pisces Overdensity, suggesting
that the Pisces stars may have originated in the Virgo progenitor.

We appreciate useful discussions with Kathryn Johnston about the
expected properties of ``cloud-like'' debris structures, and thank the
referee for thoughtful comments that improved the
manuscript. J.~L.~C. and H.~J.~N. gratefully acknowledge the support
of National Science Foundation grant AST 09-37523.

{\it Facilities:} \facility{WIYN (Hydra)}, \facility{Blanco (Hydra)},
\facility{Sloan}, \facility{MtW:1.5m}, \facility{Du Pont}

Funding for the SDSS and SDSS-II has been provided by the Alfred
P. Sloan Foundation, the Participating Institutions, the National
Science Foundation, the U.S. Department of Energy, the National
Aeronautics and Space Administration, the Japanese Monbukagakusho, the
Max Planck Society, and the Higher Education Funding Council for
England. The SDSS Web Site is http://www.sdss.org/.

The SDSS is managed by the Astrophysical Research Consortium for the
Participating Institutions. The Participating Institutions are the
American Museum of Natural History, Astrophysical Institute Potsdam,
University of Basel, University of Cambridge, Case Western Reserve
University, University of Chicago, Drexel University, Fermilab, the
Institute for Advanced Study, the Japan Participation Group, Johns
Hopkins University, the Joint Institute for Nuclear Astrophysics, the
Kavli Institute for Particle Astrophysics and Cosmology, the Korean
Scientist Group, the Chinese Academy of Sciences (LAMOST), Los Alamos
National Laboratory, the Max-Planck-Institute for Astronomy (MPIA),
the Max-Planck-Institute for Astrophysics (MPA), New Mexico State
University, Ohio State University, University of Pittsburgh,
University of Portsmouth, Princeton University, the United States
Naval Observatory, and the University of Washington.

\bibliographystyle{apj}

\begin{thebibliography}{77}
\expandafter\ifx\csname natexlab\endcsname\relax\def\natexlab#1{#1}\fi

\bibitem[{{Abadi} {et~al.}(2003){Abadi}, {Navarro}, {Steinmetz}, \&
  {Eke}}]{ans+03a}
{Abadi}, M.~G., {Navarro}, J.~F., {Steinmetz}, M., \& {Eke}, V.~R. 2003, \apj,
  591, 499

\bibitem[{{An} {et~al.}(2008){An}, {Johnson}, {Clem}, {Yanny}, {Rockosi},
  {Morrison}, {Harding}, {Gunn}, {Allende Prieto}, {Beers}, {Cudworth},
  {Ivans}, {Ivezi{\'c}}, {Lee}, {Lupton}, {Bizyaev}, {Brewington},
  {Malanushenko}, {Malanushenko}, {Oravetz}, {Pan}, {Simmons}, {Snedden},
  {Watters}, \& {York}}]{ajc+08}
{An}, D., {Johnson}, J.~A., {Clem}, J.~L., {et~al.} 2008, \apjs, 179, 326

\bibitem[{{An} {et~al.}(2009){An}, {Johnson}, {Beers}, {Pinsonneault},
  {Terndrup}, {Delahaye}, {Lee}, {Masseron}, \& {Yanny}}]{ajb+09}
{An}, D., {Johnson}, J.~A., {Beers}, T.~C., {et~al.} 2009, \apjl, 707, L64

\bibitem[{{Belokurov} {et~al.}(2007){Belokurov}, {Zucker}, {Evans}, {Kleyna},
  {Koposov}, {Hodgkin}, {Irwin}, {Gilmore}, {Wilkinson}, {Fellhauer},
  {Bramich}, {Hewett}, {Vidrih}, {De Jong}, {Smith}, {Rix}, {Bell}, {Wyse},
  {Newberg}, {Mayeur}, {Yanny}, {Rockosi}, {Gnedin}, {Schneider}, {Beers},
  {Barentine}, {Brewington}, {Brinkmann}, {Harvanek}, {Kleinman}, {Krzesinski},
  {Long}, {Nitta}, \& {Snedden}}]{bze+07}
{Belokurov}, V., {Zucker}, D.~B., {Evans}, N.~W., {et~al.} 2007, \apj, 654, 897

\bibitem[{{Brink} {et~al.}(2010){Brink}, {Mateo}, \&
  {Mart{\'{\i}}nez-Delgado}}]{bmm10}
{Brink}, T.~G., {Mateo}, M., \& {Mart{\'{\i}}nez-Delgado}, D. 2010, \aj, 140,
  1337

\bibitem[{{Bullock} \& {Johnston}(2005)}]{bj05}
{Bullock}, J.~S., \& {Johnston}, K.~V. 2005, \apj, 635, 931

\bibitem[{{Carlin} {et~al.}(2012){Carlin}, {Majewski}, {Casetti-Dinescu},
  {Law}, {Girard}, \& {Patterson}}]{cmc+12}
{Carlin}, J.~L., {Majewski}, S.~R., {Casetti-Dinescu}, D.~I., {Law}, D.~R.,
  {Girard}, T.~M., \& {Patterson}, R.~J. 2012, \apj, 744, 25

\bibitem[{{Casetti-Dinescu} {et~al.}(2009){Casetti-Dinescu}, {Girard},
  {Majewski}, {Vivas}, {Wilhelm}, {Carlin}, {Beers}, \& {van Altena}}]{cgm+09a}
{Casetti-Dinescu}, D.~I., {Girard}, T.~M., {Majewski}, S.~R., {Vivas}, A.~K.,
  {Wilhelm}, R., {Carlin}, J.~L., {Beers}, T.~C., \& {van Altena}, W.~F. 2009,
  \apjl, 701, L29
  
\bibitem[{{Casetti-Dinescu} {et~al.}(2006){Casetti-Dinescu}, {Majewski},
  {Girard}, {Carlin}, {van Altena}, {Patterson}, \& {Law}}]{cmg+06}
{Casetti-Dinescu}, D.~I., {Majewski}, S.~R., {Girard}, T.~M., {Carlin}, J.~L.,
  {van Altena}, W.~F., {Patterson}, R.~J., \& {Law}, D.~R. 2006, \aj, 132, 2082

\bibitem[{{Casey} {et~al.}(2012){Casey}, {Keller}, \& {Da Costa}}]{ckd12}
{Casey}, A.~R., {Keller}, S.~C., \& {Da Costa}, G. 2012, \aj, 143, 88

\bibitem[{{Chonis} {et~al.}(2011){Chonis}, {Mart{\'{\i}}nez-Delgado}, {Gabany},
  {Majewski}, {Hill}, {Gralak}, \& {Trujillo}}]{cmg+11}
{Chonis}, T.~S., {Mart{\'{\i}}nez-Delgado}, D., {Gabany}, R.~J., {Majewski},
  S.~R., {Hill}, G.~J., {Gralak}, R., \& {Trujillo}, I. 2011, \aj, 142, 166

\bibitem[{{Chou} {et~al.}(2007){Chou}, {Majewski}, {Cunha}, {Smith},
  {Patterson}, {Mart{\'{\i}}nez-Delgado}, {Law}, {Crane}, {Mu{\~n}oz}, {Garcia
  L{\'o}pez}, {Geisler}, \& {Skrutskie}}]{cmc+07}
{Chou}, M., {Majewski}, S.~R., {Cunha}, K., {et~al.} 2007, \apj, 670, 346

\bibitem[{{Cohen} {et~al.}(2011){Cohen}, {Huang}, \& {Kirby}}]{chk11}
{Cohen}, J.~G., {Huang}, W., \& {Kirby}, E.~N. 2011, \apj, 740, 60

\bibitem[{{Cohen} {et~al.}(2010){Cohen}, {Kirby}, {Simon}, \& {Geha}}]{cks+10}
{Cohen}, J.~G., {Kirby}, E.~N., {Simon}, J.~D., \& {Geha}, M. 2010, \apj, 725,
  288

\bibitem[{{Dehnen}(2002)}]{d02}
{Dehnen}, W. 2002, Journal of Computational Physics, 179, 27

\bibitem[{{Dellomo} \& {Newberg}(2009)}]{dn09} {Dellomo}, J., \& {Newberg}, H.~J. 2009, Bulletin of the American Astronomical Society, 41, \#425.11 

\bibitem[{{Duffau} {et~al.}(2010){Duffau}, {Vivas}, {Zinn}, {M{\'e}ndez}, \&
  {Ruiz}}]{dvz+10}
{Duffau}, S., {Vivas}, A.~K., {Zinn}, R., {M{\'e}ndez}, R.~A., \& {Ruiz}, M.~T.
  2010, in IAU Symposium, Vol. 262, IAU Symposium, ed. {G.~Bruzual \&
  S.~Charlot}, 131--134

\bibitem[{{Duffau} {et~al.}(2006){Duffau}, {Zinn}, {Vivas}, {Carraro},
  {M{\'e}ndez}, {Winnick}, \& {Gallart}}]{dzv+06}
{Duffau}, S., {Zinn}, R., {Vivas}, A.~K., {Carraro}, G., {M{\'e}ndez}, R.~A.,
  {Winnick}, R., \& {Gallart}, C. 2006, \apjl, 636, L97

\bibitem[{{Font} {et~al.}(2006){Font}, {Johnston}, {Bullock}, \&
  {Robertson}}]{fjb+06a}
{Font}, A.~S., {Johnston}, K.~V., {Bullock}, J.~S., \& {Robertson}, B.~E. 2006,
  \apj, 646, 886

\bibitem[{{Friel}(1987)}]{f87}
{Friel}, E.~D. 1987, \aj, 93, 1388

\bibitem[{{Geha} {et~al.}(2009){Geha}, {Willman}, {Simon}, {Strigari}, {Kirby},
  {Law}, \& {Strader}}]{gws+09}
{Geha}, M., {Willman}, B., {Simon}, J.~D., {Strigari}, L.~E., {Kirby}, E.~N.,
  {Law}, D.~R., \& {Strader}, J. 2009, \apj, 692, 1464

\bibitem[{{Girardi} {et~al.}(2004){Girardi}, {Grebel}, {Odenkirchen}, \&
  {Chiosi}}]{ggo+04}
{Girardi}, L., {Grebel}, E.~K., {Odenkirchen}, M., \& {Chiosi}, C. 2004, \aap,
  422, 205

\bibitem[{{Grillmair}(2010)}]{g10}
{Grillmair}, C.~J. 2010, in Galaxies and their Masks, ed. {D.~L.~Block,
  K.~C.~Freeman, \& I.~Puerari}, 247

\bibitem[{{Johnson} \& {Soderblom}(1987)}]{js87}
{Johnson}, D.~R.~H., \& {Soderblom}, D.~R. 1987, \aj, 93, 864

\bibitem[{{Johnston}(1998)}]{j98a}
{Johnston}, K.~V. 1998, \apj, 495, 297

\bibitem[{{Johnston} {et~al.}(2008){Johnston}, {Bullock}, {Sharma}, {Font},
  {Robertson}, \& {Leitner}}]{jbs+08}
{Johnston}, K.~V., {Bullock}, J.~S., {Sharma}, S., {Font}, A., {Robertson},
  B.~E., \& {Leitner}, S.~N. 2008, \apj, 689, 936

\bibitem[{{Jones}(1998)}]{j98}
{Jones}, L.~A. 1998, PhD thesis, The University of North Carolina at Chapel
  Hill

\bibitem[{{Juri{\'c}} {et~al.}(2008){Juri{\'c}}, {Ivezi{\'c}}, {Brooks},
  {Lupton}, {Schlegel}, {Finkbeiner}, {Padmanabhan}, {Bond}, {Sesar},
  {Rockosi}, {Knapp}, {Gunn}, {Sumi}, {Schneider}, {Barentine}, {Brewington},
  {Brinkmann}, {Fukugita}, {Harvanek}, {Kleinman}, {Krzesinski}, {Long},
  {Neilsen}, {Nitta}, {Snedden}, \& {York}}]{jib+08}
{Juri{\'c}}, M., {Ivezi{\'c}}, {\v Z}, {Brooks}, A., {et~al.} 2008, \apj, 673, 864

\bibitem[{{Kapteyn}(1906)}]{k06}
{Kapteyn}, J.~C. 1906, {Plan of selected areas} (Groningen, Hoitsema brothers,
  1906.)

\bibitem[{{Keller}(2010)}]{k10a}
{Keller}, S.~C. 2010, \pasa, 27, 45

\bibitem[{{Keller} {et~al.}(2009){Keller}, {da Costa}, \& {Prior}}]{kdp09}
{Keller}, S.~C., {da Costa}, G.~S., \& {Prior}, S.~L. 2009, \mnras, 394, 1045

\bibitem[{{Keller} {et~al.}(2008){Keller}, {Murphy}, {Prior}, {Da Costa}, \&
  {Schmidt}}]{kmp+08}
{Keller}, S.~C., {Murphy}, S., {Prior}, S., {Da Costa}, G., \& {Schmidt}, B.
  2008, \apj, 678, 851

\bibitem[{{Kollmeier} {et~al.}(2009){Kollmeier}, {Gould}, {Shectman},
  {Thompson}, {Preston}, {Simon}, {Crane}, {Ivezi{\'c}}, \& {Sesar}}]{kgs+09}
{Kollmeier}, J.~A., {Gould}, A., {Shectman}, S., {et~al.} 2009, \apjl, 705, L158

\bibitem[{{Law} {et~al.}(2005){Law}, {Johnston}, \& {Majewski}}]{ljm05}
{Law}, D.~R., {Johnston}, K.~V., \& {Majewski}, S.~R. 2005, \apj, 619, 807

\bibitem[{{Law} \& {Majewski}(2010)}]{lm10a}
{Law}, D.~R., \& {Majewski}, S.~R. 2010, \apj, 714, 229

\bibitem[{{Mackey} \& {van den Bergh}(2005)}]{mv05}
{Mackey}, A.~D., \& {van den Bergh}, S. 2005, \mnras, 360, 631

\bibitem[{{Majewski}(1993)}]{m93}
{Majewski}, S.~R. 1993, \araa, 31, 575

\bibitem[{{Majewski} {et~al.}(1996){Majewski}, {Munn}, \& {Hawley}}]{mmh96}
{Majewski}, S.~R., {Munn}, J.~A., \& {Hawley}, S.~L. 1996, \apjl, 459, L73+

\bibitem[{{Majewski} {et~al.}(2003){Majewski}, {Skrutskie}, {Weinberg}, \&
  {Ostheimer}}]{msw+03}
{Majewski}, S.~R., {Skrutskie}, M.~F., {Weinberg}, M.~D., \& {Ostheimer}, J.~C.
  2003, \apj, 599, 1082

\bibitem[{{Mart{\'{\i}}nez-Delgado} {et~al.}(2004){Mart{\'{\i}}nez-Delgado},
  {G{\'o}mez-Flechoso}, {Aparicio}, \& {Carrera}}]{mga+04}
{Mart{\'{\i}}nez-Delgado}, D., {G{\'o}mez-Flechoso}, M.~{\'A}., {Aparicio}, A.,
  \& {Carrera}, R. 2004, \apj, 601, 242

\bibitem[{{Mart{\'{\i}}nez-Delgado} {et~al.}(2010){Mart{\'{\i}}nez-Delgado},
  {Gabany}, {Crawford}, {Zibetti}, {Majewski}, {Rix}, {Fliri},
  {Carballo-Bello}, {Bardalez-Gagliuffi}, {Pe{\~n}arrubia}, {Chonis}, {Madore},
  {Trujillo}, {Schirmer}, \& {McDavid}}]{mgc+10}
{Mart{\'{\i}}nez-Delgado}, D., {Gabany}, R.~J., {Crawford}, K., {et~al.} 2010, \aj, 140, 962

\bibitem[{{Miyamoto} \& {Nagai}(1975)}]{mn75}
{Miyamoto}, M., \& {Nagai}, R. 1975, \pasj, 27, 533

\bibitem[{{Mu{\~n}oz} {et~al.}(2006){Mu{\~n}oz}, {Carlin}, {Frinchaboy},
  {Nidever}, {Majewski}, \& {Patterson}}]{mcf+06}
{Mu{\~n}oz}, R.~R., {Carlin}, J.~L., {Frinchaboy}, P.~M., {Nidever}, D.~L.,
  {Majewski}, S.~R., \& {Patterson}, R.~J. 2006, \apjl, 650, L51

\bibitem[{{Newberg} {et~al.}(2010){Newberg}, {Willett}, {Yanny}, \&
  {Xu}}]{nwy+10}
{Newberg}, H.~J., {Willett}, B.~A., {Yanny}, B., \& {Xu}, Y. 2010, \apj, 711,
  32

\bibitem[{{Newberg} {et~al.}(2007){Newberg}, {Yanny}, {Cole}, {Beers}, {Re
  Fiorentin}, {Schneider}, \& {Wilhelm}}]{nyc+07}
{Newberg}, H.~J., {Yanny}, B., {Cole}, N., {Beers}, T.~C., {Re Fiorentin}, P.,
  {Schneider}, D.~P., \& {Wilhelm}, R. 2007, \apj, 668, 221

\bibitem[{{Newberg} {et~al.}(2002){Newberg}, {Yanny}, {Rockosi}, {Grebel},
  {Rix}, {Brinkmann}, {Csabai}, {Hennessy}, {Hindsley}, {Ibata}, {Ivezi{\'c}},
  {Lamb}, {Nash}, {Odenkirchen}, {Rave}, {Schneider}, {Smith}, {Stolte}, \&
  {York}}]{nyr+02}
{Newberg}, H.~J., {Yanny}, B., {Rockosi}, C., {et~al.} 2002, \apj, 569, 245

\bibitem[{{Newberg} {et~al.}(2003){Newberg}, {Yanny}, {Grebel}, {Hennessy},
  {Ivezi{\'c}}, {Martinez-Delgado}, {Odenkirchen}, {Rix}, {Brinkmann}, {Lamb},
  {Schneider}, \& {York}}]{nyg+03}
---. 2003, \apjl, 596, L191

\bibitem[{{Norris} {et~al.}(2010){Norris}, {Wyse}, {Gilmore}, {Yong}, {Frebel},
  {Wilkinson}, {Belokurov}, \& {Zucker}}]{nwg+10}
{Norris}, J.~E., {Wyse}, R.~F.~G., {Gilmore}, G., {et al.} 2010, \apj, 723, 1632 

\bibitem[{{Peterson} {et~al.}(1986){Peterson}, {Olszewski}, \&
  {Aaronson}}]{poa86}
{Peterson}, R.~C., {Olszewski}, E.~W., \& {Aaronson}, M. 1986, \apj, 307, 139

\bibitem[{{Prior} {et~al.}(2009){Prior}, {Da Costa}, {Keller}, \&
  {Murphy}}]{pdk+09}
{Prior}, S.~L., {Da Costa}, G.~S., {Keller}, S.~C., \& {Murphy}, S.~J. 2009,
  \apj, 691, 306

\bibitem[{{Robin} {et~al.}(2003){Robin}, {Reyl{\'e}}, {Derri{\`e}re}, \&
  {Picaud}}]{rrd+03}
{Robin}, A.~C., {Reyl{\'e}}, C., {Derri{\`e}re}, S., \& {Picaud}, S. 2003,
  \aap, 409, 523

\bibitem[{{Rocha-Pinto}(2010)}]{r10a}
{Rocha-Pinto}, H.~J. 2010, in IAU Symposium, Vol. 265, IAU Symposium, ed.
  {K.~Cunha, M.~Spite, \& B.~Barbuy}, 255--262

\bibitem[{{Schiavon}(2007)}]{s07}
{Schiavon}, R.~P. 2007, \apjs, 171, 146

\bibitem[{{Schlegel} {et~al.}(1998){Schlegel}, {Finkbeiner}, \&
  {Davis}}]{sfd98}
{Schlegel}, D.~J., {Finkbeiner}, D.~P., \& {Davis}, M. 1998, \apj, 500, 525

\bibitem[{{Seares} {et~al.}(1930){Seares}, {Kapteyn}, {van Rhijn}, {Joyner}, \&
  {Richmond}}]{skv+30}
{Seares}, F.~H., {Kapteyn}, J.~C., {van Rhijn}, P.~J., {Joyner}, M.~C., \&
  {Richmond}, M.~L. 1930, {Mount Wilson catalogue of photographic magnitudes in
  selected areas 1-139} (Carnegie institution of Washington)

\bibitem[{{Searle} \& {Zinn}(1978)}]{sz78}
{Searle}, L., \& {Zinn}, R. 1978, \apj, 225, 357

\bibitem[{{Sesar} {et~al.}(2010{\natexlab{a}}){Sesar}, {Vivas}, {Duffau}, \&
  {Ivezi{\'c}}}]{svd+10}
{Sesar}, B., {Vivas}, A.~K., {Duffau}, S., \& {Ivezi{\'c}}, {\v Z}.
  2010{\natexlab{a}}, \apj, 717, 133

\bibitem[{{Sesar} {et~al.}(2007){Sesar}, {Ivezi{\'c}}, {Lupton}, {Juri{\'c}},
  {Gunn}, {Knapp}, {DeLee}, {Smith}, {Miknaitis}, {Lin}, {Tucker}, {Doi},
  {Tanaka}, {Fukugita}, {Holtzman}, {Kent}, {Yanny}, {Schlegel}, {Finkbeiner},
  {Padmanabhan}, {Rockosi}, {Bond}, {Lee}, {Stoughton}, {Jester}, {Harris},
  {Harding}, {Brinkmann}, {Schneider}, {York}, {Richmond}, \& {Vanden
  Berk}}]{sil+07}
{Sesar}, B., {Ivezi{\'c}}, {\v Z}, {Lupton}, R.~H., {et~al.} 2007, \aj, 134, 2236

\bibitem[{{Sesar} {et~al.}(2010{\natexlab{b}}){Sesar}, {Ivezi{\'c}}, {Grammer},
  {Morgan}, {Becker}, {Juri{\'c}}, {De Lee}, {Annis}, {Beers}, {Fan}, {Lupton},
  {Gunn}, {Knapp}, {Jiang}, {Jester}, {Johnston}, \& {Lampeitl}}]{sig+10}
---. 2010{\natexlab{b}}, \apj, 708, 717

\bibitem[{{Sharma} {et~al.}(2010){Sharma}, {Johnston}, {Majewski}, {Mu{\~n}oz},
  {Carlberg}, \& {Bullock}}]{sjm+10}
{Sharma}, S., {Johnston}, K.~V., {Majewski}, S.~R., {Mu{\~n}oz}, R.~R.,
  {Carlberg}, J.~K., \& {Bullock}, J. 2010, \apj, 722, 750

\bibitem[{{Simon} \& {Geha}(2007)}]{sg07}
{Simon}, J.~D., \& {Geha}, M. 2007, \apj, 670, 313

\bibitem[{{Simon} {et~al.}(2011){Simon}, {Geha}, {Minor}, {Martinez}, {Kirby},
  {Bullock}, {Kaplinghat}, {Strigari}, {Willman}, {Choi}, {Tollerud}, \&
  {Wolf}}]{sgm+11}
{Simon}, J.~D.,  {Geha}, M., {Minor}, Q.~E., {et~al.} 2011, \apj, 733, 46

\bibitem[{{Suntzeff} {et~al.}(1988){Suntzeff}, {Kraft}, \& {Kinman}}]{skk88}
{Suntzeff}, N.~B., {Kraft}, R.~P., \& {Kinman}, T.~D. 1988, \aj, 95, 91

\bibitem[{{Teuben}(1995)}]{t95}
{Teuben}, P. 1995, in Astronomical Society of the Pacific Conference Series,
  Vol.~77, Astronomical Data Analysis Software and Systems IV, ed. {R.~A.~Shaw,
  H.~E.~Payne, \& J.~J.~E.~Hayes}, 398--+

\bibitem[{{Tonry} \& {Davis}(1979)}]{td79}
{Tonry}, J., \& {Davis}, M. 1979, \aj, 84, 1511

\bibitem[{{van den Bergh}(2007)}]{v07a}
{van den Bergh}, S. 2007, \aj, 134, 344

\bibitem[{{van den Bergh} \& {Mackey}(2004)}]{vm04}
{van den Bergh}, S., \& {Mackey}, A.~D. 2004, \mnras, 354, 713

\bibitem[{{Vivas} {et~al.}(2008){Vivas}, {Jaff{\'e}}, {Zinn}, {Winnick},
  {Duffau}, \& {Mateu}}]{vjz+08}
{Vivas}, A.~K., {Jaff{\'e}}, Y.~L., {Zinn}, R., {Winnick}, R., {Duffau}, S., \&
  {Mateu}, C. 2008, \aj, 136, 1645

\bibitem[{{Vivas} \& {Zinn}(2003)}]{vz03}
{Vivas}, A.~K., \& {Zinn}, R. 2003, \memsai, 74, 928

\bibitem[{{Vivas} \& {Zinn}(2006)}]{vz06}
---. 2006, \aj, 132, 714

\bibitem[{{Vivas} {et~al.}(2001){Vivas}, {Zinn}, {Andrews}, {Bailyn}, {Baltay},
  {Coppi}, {Ellman}, {Girard}, {Rabinowitz}, {Schaefer}, {Shin}, {Snyder},
  {Sofia}, {van Altena}, {Abad}, {Bongiovanni}, {Brice{\~n}o}, {Bruzual},
  {Della Prugna}, {Herrera}, {Magris}, {Mateu}, {Pacheco}, {S{\'a}nchez},
  {S{\'a}nchez}, {Schenner}, {Stock}, {Vicente}, {Vieira}, {Ferr{\'{\i}}n},
  {Hernandez}, {Gebhard}, {Honeycutt}, {Mufson}, {Musser}, \&
  {Rengstorf}}]{vza+01}
{Vivas}, A.~K., {Zinn}, R., {Andrews}, P., {et~al.} 2001, \apjl, 554, L33

\bibitem[{{Vivas} {et~al.}(2004){Vivas}, {Zinn}, {Abad}, {Andrews}, {Bailyn},
  {Baltay}, {Bongiovanni}, {Brice{\~n}o}, {Bruzual}, {Coppi}, {Della Prugna},
  {Ellman}, {Ferr{\'{\i}}n}, {Gebhard}, {Girard}, {Hernandez}, {Herrera},
  {Honeycutt}, {Magris}, {Mufson}, {Musser}, {Naranjo}, {Rabinowitz},
  {Rengstorf}, {Rosenzweig}, {S{\'a}nchez}, {S{\'a}nchez}, {Schaefer},
  {Schenner}, {Snyder}, {Sofia}, {Stock}, {van Altena}, {Vicente}, \&
  {Vieira}}]{vza+04}
---. 2004, \aj, 127, 1158

\bibitem[{{Vogt} {et~al.}(1995){Vogt}, {Mateo}, {Olszewski}, \&
  {Keane}}]{vmo+95}
{Vogt}, S.~S., {Mateo}, M., {Olszewski}, E.~W., \& {Keane}, M.~J. 1995, \aj,
  109, 151

\bibitem[{{Walker} {et~al.}(2009){Walker}, {Mateo}, {Olszewski},
  {Pe{\~n}arrubia}, {Wyn Evans}, \& {Gilmore}}]{wmo+09}
{Walker}, M.~G., {Mateo}, M., {Olszewski}, E.~W., {Pe{\~n}arrubia}, J., {Wyn
  Evans}, N., \& {Gilmore}, G. 2009, \apj, 704, 1274

\bibitem[{{Watkins} {et~al.}(2009){Watkins}, {Evans}, {Belokurov}, {Smith},
  {Hewett}, {Bramich}, {Gilmore}, {Irwin}, {Vidrih}, {Wyrzykowski}, \&
  {Zucker}}]{web+09}
{Watkins}, L.~L., {Evans}, N.~W., {Belokurov}, V., {et~al.} 2009, \mnras, 398, 1757

\bibitem[{{Willett} {et~al.}(2009){Willett}, {Newberg}, {Zhang}, {Yanny}, \&
  {Beers}}]{wnz+09}
{Willett}, B.~A., {Newberg}, H.~J., {Zhang}, H., {Yanny}, B., \& {Beers}, T.~C.
  2009, \apj, 697, 207

\bibitem[{{Wolf} {et~al.}(2010){Wolf}, {Martinez}, {Bullock}, {Kaplinghat},
  {Geha}, {Mu{\~n}oz}, {Simon}, \& {Avedo}}]{wmb+10}
{Wolf}, J., {Martinez}, G.~D., {Bullock}, J.~S., {Kaplinghat}, M., {Geha}, M.,
  {Mu{\~n}oz}, R.~R., {Simon}, J.~D., \& {Avedo}, F.~F. 2010, \mnras, 406, 1220

\bibitem[{{Worthey} {et~al.}(1994){Worthey}, {Faber}, {Gonzalez}, \&
  {Burstein}}]{wfg+94}
{Worthey}, G., {Faber}, S.~M., {Gonzalez}, J.~J., \& {Burstein}, D. 1994,
  \apjs, 94, 687

\end{thebibliography}

\end{document}